\documentclass[journal]{IEEEtran}
\usepackage{graphicx}
\usepackage{amsmath}
\usepackage{multirow}
\usepackage{amsfonts,amsmath}
\usepackage{amssymb, nccmath}
\usepackage{amssymb, nccmath}
\usepackage{dsfont}
\usepackage{booktabs}
\usepackage[utf8]{inputenc}
\usepackage{dsfont}
\usepackage{amsfonts}
\usepackage{booktabs}
\usepackage{graphicx}
\usepackage{grffile}
\usepackage{grffile}
\usepackage{array} 
\usepackage{mathtools}

\usepackage{subfigure} 
\usepackage{fontenc}
\usepackage{tikz}
\usepackage{pgfplots}
\usepackage{tikz}
\usetikzlibrary{patterns}
\definecolor{myhund}{HTML}{9966CC}
\definecolor{myfifty}{HTML}{002E63}
\definecolor{mytwenty}{HTML}{4997D0}
\definecolor{s1}{HTML}{66FF00}
\definecolor{s2}{HTML}{00CC99}
\definecolor{s3}{HTML}{8DB600}
\definecolor{s4}{HTML}{177245}
\definecolor{r1}{HTML}{FF004F}
\definecolor{r2}{HTML}{FF4F00 }
\definecolor{r3}{HTML}{F400A1}
\definecolor{r4}{HTML}{CE2029}
\usetikzlibrary{positioning}
\definecolor{Mycolor}{HTML}{03C03C}
\usepackage[linesnumbered, ruled]{algorithm2e}
\makeatletter
\newcommand{\removelatexerror}{\let\@latex@error\@gobble}
\def\ps@IEEEtitlepagestyle{%
	\def\@oddfoot{\mycopyrightnotice}%
	\def\@oddhead{\hbox{}\@IEEEheaderstyle\leftmark\hfil\thepage}\relax
	\def\@evenhead{\@IEEEheaderstyle\thepage\hfil\leftmark\hbox{}}\relax
	\def\@evenfoot{}%
}

\def\mycopyrightnotice{%
	\begin{minipage}{\textwidth}
		\centering \scriptsize
		© 2024 IEEE. This article has been accepted in IEEE Transactions on Services Computing Journal © 2024 IEEE. Personal use of this material is permitted. Permission from IEEE must be obtained for all other uses, in any current or future media, including reprinting/republishing this material for advertising or promotional purposes, creating new collective works, for resale or redistribution to servers or lists, or reuse of any copyrighted component of this work in other works. This work is freely available for survey and citation.		
	\end{minipage}
}
\makeatother

\ifCLASSOPTIONcompsoc
  \usepackage[nocompress]{cite}
\else
  \usepackage{cite}
\fi
\ifCLASSINFOpdf
\else
\fi
\begin{document}
%

\title{An Oversubscription and Service Pricing Exploitation-based Profit Maximization Framework for  Industry Cloud Resource Management }

\author{Deepika~Saxena,~\IEEEmembership{Member~IEEE} and
	Ashutosh~Kumar~Singh,~\IEEEmembership{Senior~Member~IEEE}
	\thanks{D. Saxena is with the School of Computer Science and Engineering, The University of Aizu, Japan. E-mail: 13deepikasaxena@gmail.com\\
	A.K. Singh is with the Department of Computer Science and Engineering, Indian Institute of Information Technology, Bhopal, India and also with Department of Computer Science, the University of Economics and Human Sciences, 01-043 Warsaw, Poland. E-mail: ashutosh@iiitbhopal.ac.in}\\
 }

\IEEEtitleabstractindextext{%
\begin{abstract}
This paper proposed a novel industry cloud resource management framework that exploits resource oversubscription and heterogeneous service pricing models to maximize profitability and operational efficiency for industry cloud providers. The framework proposes an adaptive ensemble machine learning driven prediction model for proactive estimation of resource utilization of Virtual Machines (VM)s-based on previous resource utilization of respective users' VMs to minimize resource wastage due to oversubscription by them. Accordingly, the VMs having similar predicted resource usage are grouped using Fuzzy C-means clustering. This helps   to determine the required number of VMs with specific configuration to be deployed before executing user requests. Concurrently, the framework incorporates two distinct categories of cloud service pricing models, namely the \textit{Delay Sensitive Model} and the \textit{Best-Effort Model}. Accordingly,  the user requests are classified  and executed by selecting the most suitable VMs, with the goal of maximizing revenue and reducing electricity costs in cloud data centers ($\mathds{CDC}$s). Experimental simulation and comparison against state-of-the-art methods, using two benchmark VM traces, validates the performance of  proposed framework. It significantly reduces electricity bills by 55.56\%, power consumption and active servers by up to 60.7\% and 51\%, respectively, while improving resource utilization and profits by up to 60\% and 51.18\%, respectively. 
\end{abstract}

}

\maketitle

\IEEEdisplaynontitleabstractindextext

%
\IEEEpeerreviewmaketitle

{\section{Introduction}\label{sec:introduction}}
\IEEEPARstart{I}{ndustry} cloud platforms are specialized cloud platforms that cater to the specific needs and requirements of a particular industry or vertical market. These platforms provide industry-specific functionalities, tools, and services, transforming a generic cloud platform into a business platform tailored to the needs of that industry. Power consumption represents a substantial portion of operational costs for industry cloud providers (ICP)s. 
 Large scale industrial cloud providers such as Google and Amazon pay  electricity bills in millions of dollars per month for the huge power consumption in their data centres  \cite{datacenterdynamics2021}. The estimated annual electricity consumption of a data centre vary from 200 TeraWatt Hours (TWH) to 500 TWH that accounts for 30 to 50 percent of their total operational cost \cite{kaur2019big}. Therefore, ICPs seek for an effective financial incentive to prune or minimize their operational cost in every possible way.  This cost savings can be passed on to customers, making industry cloud services more affordable and competitive.  The recent researches  \cite{saxena2021op, saxena2022high,li2019transforming} indicate that most of the time, maximal resource utilization of servers varies between 10\% and 50\% only that leads to resource wastage and excess power consumption because of being `active' in idle or under-utilized state  \cite{yuan2016ttsa}. The main reason behind this is over reservation of physical resources viz., CPU and memory by the industry cloud users during selection of VMs to allow contention-free execution of their requests \cite{singh2015qos, saxena2023ai}. The users over-estimate the resources' capacities as they are unknown to the actual demand of resources prior to execution \cite{saxena2021op}. Also, they anticipate that the requirement of resources may reach its peak and vary over time  \cite{singh2021quantum, saxena2024high} leading to inappropriate selection of cloud network. Such a concern of cloud users raises a key challenge for ICP that how to minimize the excessive operational cost due to over-reserved and under-utilized physical resources?   

\begin{figure}[!htbp]	
	\centering
	\subfigure[Power consumption ]{\includegraphics[width=.23\textwidth]{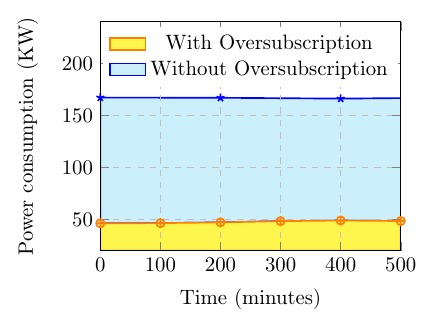}}
	\subfigure[Operational Cost ]{\includegraphics[width=.23\textwidth]{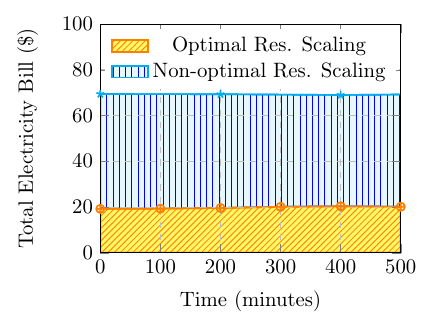}}\hfill	
	\caption{Breakdown of power consumption based Google CDC \cite{reiss2011google} electricity bill  }
	\label{fig:motivation}	
\end{figure}

An efficient implementation of oversubscription of indusrty cloud resources provides a potential solution to the aforementioned problem. The term \textit{oversubscription} in the context of industry cloud resources refers to the practice of allocating resources to VMs in excess of the actual physical resource capacity of the server hosting them. This approach is based on the assumption that not all VMs will fully utilize their allocated resources simultaneously \cite{singh2023bio}. Instead, it relies on statistical multiplexing to allow for efficient resource utilization across multiple VMs. It allows cloud providers to optimize resource utilization, minimize power consumption, and maximize efficiency, leading to cost savings and improved scalability. With oversubscription  service model, ICPs anticipate that users' VMs are not likely to utilize their actual demanded resources fully, and allocate resources to VMs in excess of the actual resource capacity of the server hosting them \cite{son2017sla, molto2016automatic}. Furthermore, this fact is analysed and computed by conducting an experimental measurement on Google Cluster VM traces \cite{reiss2011google} as depicted in Fig. \ref{fig:motivation}.
The comparison of power consumption of a VM with oversubscription and without oversubscription for duration of 500 minutes is reported in Fig. \ref{fig:motivation}(a) and its associated electricity bill is shown in Fig. \ref{fig:motivation}(b). The resource utilization are parsed from the VM traces and power consumption is evaluated by applying power model given in \cite{saxena2021op}, \cite{sharma2016multi}, \cite{minas2009energy}: $PW_{dc} = \sum_{i=1}^{P} {[{PW_i}^{max} - {PW_i}^{min}]{RU} + {PW_i}^{idle}}$, where ${PW_i}^{max}$=135 Watts, ${PW_i}^{min}$= ${PW_i}^{idle}$=93.7 Watts \cite{Dell1999}, and $RU$ is resource utilization for data centre of size =1200 VMs. The electricity bill is computed using real power prices, where energy charge = $0.07\$/KWH$ \cite{usEnergy2014}. It can be clearly observed that optimal resource scaling achieved with oversubscription has reduced power consumption and operational cost up to $69.4\%$  against without oversubscription resource allocation. Therefore, oversubscription leverages a great opportunity for minimizing power consumption and operational cost in industrial $\mathds{CDC}$s by consolidating entire workload on fewer number of active servers. 

\par Furthermore, industry clouds offer variety of service pricing models which have a huge influence on ICPs' earned revenues. By exploiting the heterogeneity of industry cloud service pricing models and  practices, ICPs can significantly reduce their energy bills. Inspite of numerous benefits of oversubscription service model and exploitation of heterogeneous service pricing models, ICP restricts their execution in the data centres because of the challenges associated to its efficient implementation including heterogeneous workload, uncertain resource demands, live VM migration and downtime etc. Cloud users submit different type of workloads, requiring heterogeneous resource capacities with varying priorities and pricing policies with varying resource demands over time making it difficult to decide resource distribution \cite{pinciroli2020cedule+}.

\subsection{Paper Contributions}
This paper proposes an \textbf{O}nline \textbf{P}roactive Resource Scaling \textbf{F}ramework for \textbf{P}rofit \textbf{M}aximization (\textbf{OP-PMF}) for Industry Cloud Providers  that addresses all the aforementioned challenges and optimizes operational cost of industrial $\mathds{CDC}$s by exploiting and leveraging oversubsciption model and service pricing models. Concerning oversubsciption model, the future resource utilization of users' VMs are proactively estimated via proposed adaptive ensemble learning based online VM predictor which regenerates periodically respective to achieved accuracy score of different VMs. Thereafter, the predicted VMs are grouped according to the similarity of their resource usage by applying concepts of Fuzzy C-means clustering. The number of VMs are determined by the total number of elements in a particular cluster, while maximum resource capacity VM in each cluster decides the desired size or type of each VM in that cluster. These two steps help to consolidate upcoming requests on the minimum number of active servers leading to optimize resource utilization and power consumption. Further, the cloud service pricing models are utilized by classifying users' requests  into two service pricing models including \textit{Delay Sensitive Model} and \textit{Best-Effort Model} to maximize the profits for industry clouds.

{The key contributions are summarised as follows:}
\begin{itemize}
	\item OP-PMF maximizes profits of the industry $\mathds{CDC}$  by exploiting oversubscription and heterogeneous service pricing models. It minimizes the operational cost by controlling power consumption during physical resource  management and leveraging interactions
	among online resource prediction, VM autoscaling, and cost-effective load distribution.
	
	\item An online and adaptive ensemble approach based VM predictor is developed which is inspired from "No one fits All" strategy. It generates/re-generates a separate ensemble model periodically to predict future resource usage of each VM accurately.
	
	\item Self-adaptive and proactive resource scaling by clustering of predicted VMs on the basis of similarity of predicted resource usage and considering service pricing models is proposed. It helps to determine exact number and size of VMs required to execute upcoming workload.
	
	\item {Implementation and evaluation of proposed framework by using two distinct real benchmark datasets reveal that OP-PMF outperforms the state-of-arts in terms of performance metrics like VM prediction, resource utilization,  number of active servers, power saving, and electricity bill optimization.}
\end{itemize}
\par \textit{Paper organization}: Section II discusses related work. The profit maximization problem is  formulated in Section III. The proposed OP-PMF design flow  is presented in Section \ref{sec-proposed} followed by the detailed description of its sub-units including: ensemble learning-based resource prediction, leveraging cloud oversubscription, and service pricing models exploitation are conferred in Section \ref{sec-adaptive resource prediction}, Section \ref{sec-leveraging}, and Section \ref{sec-Request Execution}, respectively. Section \ref{sec-operational design} entails operational design and complexity computation. Section \ref{sec-performance} evaluates and compares performance of OP-PMF which is followed by conclusive remarks and future scope of the proposed work in Section \ref{sec-conclusion}.  Table \ref{table:notation} shows the list of symbols with their explanatory terms used throughout  this paper. 
\begin{table}[htbp]
	\centering	
	\caption[Table caption text] {Notations and their descriptions}  
	\label{table:notation}
	\resizebox{0.50\textwidth}{!}{
		\begin{tabular}{|l|}
			\hline
			{$U$}{: user;} {$M$: number of users;}
			{$V$: VM ;}  {$Q$: number of VMs; }\\  \hline
			$S$: server; $S^{\ddagger}$: number of active servers; {$P$: number of servers;}  \\ \hline    
			
			{$i$: index for servers;} {$j$: index for VMs;} {$k$: index for User;} \\ \hline {$\omega_{kji}$: Mapping of $k^{th}$ user's $j^{th}$ VM on $i^{th}$ server;}\\
			\hline
			{$RU$: resource utilization;  $PW$: power consumption;} \\ 				\hline $r$: user's request; $r^{DSM}$: DSM request; $r^{BEM}$: BEM request; \\
			\hline 
			{$R$: resources; $C$: CPU; $M$: Memory; $\mathds{N}$: Number of resources}\\ \hline
			{$\tau_{k}$: execution time of $k^{th}$ request;} {$\mathds{T}_k$: Deadline of $k^{th}$ request;}\\ \hline
			{$Z_{pr}$: predicted output; $\mathds{C}$: clusters; $V^{C}$: VM cluster;}\\ \hline
			$\xi$: mean squared error; $\hat{\xi}$: normalised error;  \\ \hline $X$: Number of base learners;
			$t$: variable for time-interval;\\ \hline
			{$WS$: weighted score;} {$U^\ddagger$: user pays more charges;} \\ \hline
			
			\hline
		\end{tabular}
	}
\end{table}

\section{Related Work}
An energy-efficient and online resource prediction and multi-objective load-balancing (OP-MLB) framework is proposed in \cite{saxena2021op}. A neural-network based prediction system is developed and employed on each VM to estimate the future resource utilization of the servers proactively and balance the load accordingly.   Yu et al. \cite{yuan2016ttsa} addressed the cost minimization problem in hybrid clouds by proposing temporal task scheduling algorithm (TTSA). It dispatches all incoming tasks to the appropriate hosts, where  the cost minimization problem is formulated as a mixed integer linear program during each iteration. This problem is solved by combining simulated annealing and particle swarm optimization techniques. For VM placement, Jangiti et al. \cite{jangiti2019aggregated} presented the First-Fit approach, Jung et al. \cite{jung2010mistral} have utilized Random-Fit based VM scheduling, and Shrivastava et al. \cite{shirvastava2017best} devised Best-Fit heuristics.
The cloud task elasticity and price heterogeneity are exploited in \cite{dabbagh2015exploiting} to propose an online resource management framework that maximizes cloud profits while minimizing energy expenses. This is done by reducing the duration during which servers need to be left on and maximizing the monetary revenues when the charging cost for some of the elastic tasks depends on how fast these tasks complete. 
Nguyen et al. \cite{hieu2017virtual} addressed the VM consolidation problem
by adopting  overloaded host detection with multiple usage prediction (OHD-MUP) and underloaded host detection with multiple usage prediction (UHD-MUP) and balanced load by migrating selected VMs from overloaded servers to energy-efficient server. When there is no overutilized host, the underloaded server migration starts and the host with least value of maximum resource utilization is considered to consolidate the load on minimum number of active servers.

\par 
Li et al. \cite{li2019transforming} addressed the challenge of reducing cooling energy costs in data centers (DCs) while ensuring thermal safety. The work proposed an end-to-end cooling control algorithm based on deep reinforcement learning (DRL) to optimize data center operations. 
Alanazi et al. \cite{alanazi2017reducing} proposed a distributive UPS topology for cloud resource management at the server and rack levels. It manages VM placement, battery charging and discharging times, and battery selection to minimize peak demands and electricity costs. VM requests are scheduled using a Slack and Battery Aware (SBA) placement, considering server power states, resource utilization, available capacity, and stored energy.    Kumar et al. \cite{kumar2018renewable} proposed a Container-as-a-service (CoaaS) model to to sustain the energy consumption of cloud data centres using renewable energy sources. An energy-aware multi-indexed job classification and scheduling
approach is designed. This scheme allocates incoming workload to geographically distributed datacentres which have sufficient amount of renewable energy to handle the incoming jobs.
Sahoo et al. \cite{sahoo2017lvrm} proposed a Link Based Virtual Resource Management (LVRM) algorithm to reduce the number of active physical machines. This algorithm maps virtual links and nodes to minimize their impact on execution time, prioritizing virtual links with maximum network bandwidth to expedite task execution. It also consolidates multiple VMs onto a single physical machine and uses Dijkstra's algorithm to select the optimal substrate path between physical machines, enhancing the request execution rate.

\textit{Research gaps and motivation for proposed framework }: The existing approaches to industry cloud resource management primarily focus on load balancing and task scheduling for energy efficiency, overlooking the potential benefits of leveraging oversubscription and heterogeneous service pricing models to maximize profits for cloud providers  while optimizing power consumption. Thus, there is a significant research gap in addressing these aspects comprehensively. In this context, a novel OP-PMF framework is proposed to fill this gap by integrating resource management operations, exploiting service pricing models and oversubscription, and optimizing costs and profits while ensuring quality of service and minimizing resource and electricity expenditure.

\section{Problem Formulation}

Consider $m$ cloud users \{${U_1,U_2,...,U_m}$\}  have purchased VMs \{$ {V_1,V_2,...,V_q}$\}  of heterogeneous configuration in a \textit{Cloud Data Center} ($\mathds{CDC}$). Let $k^{th}$ user $U_k$ submits request $r_k:\{{R}_k, t^{start}_k,  t^{end}_k \}$, where ${R}_k$ represents demanded capacity of resources: CPU (${C}_k$) and memory (${M}_k$) by the $k^{th}$ user such as ${R}_k:\{{C}_k, {M}_k\}$ for duration [\textit{start time} ($t^{start}_k$), \textit{end time} ($t^{end}_k$)]. The $k^{th}$ user's request ($r_k$) is executed in $j^{th}$ VM such that \{$V_{jk}: \forall j \in [1, q], k \in [1, m]$\} deployed on $p$ servers \{$ {S_1, S_2,...,S_p}$\}  configured with varying resources capacities. 

Specifically, the problem is how to decide the physical resource distribution \{$S_1$, $S_2$, ..., $S_p$\} among users' VMs \{$V_1$, $V_2$, ..., $V_q$\} for maximising the profit earnings ($\mathds{PE}$) for ICPs while serving  each user's request $r_k:\{{R}_k, t^{start}_k,  t^{end}_k \}| \forall_{k} \in [1,m]$ and avoiding any SLA term violation in an oversubscribed cloud environment?

\par  Let physical resource distribution for user request execution is represented via mapping: \{$\omega_{kji}|\omega_{kji}:U_k \times V_j \times S_i$\} which specifies placement of VM $V_j$ of user $U_k$ on server $S_i$ for a time interval \{$t_1$, $t_2$\}. The essential set of constraints that must be contented concurrently have been formulated in Eqs. (\ref{1})-(\ref{7}):

\begin{equation}
\textbf{$C_1$:} \quad \sum_{k \in m}\sum_{j \in q}\sum_{i \in p}{\omega_{kji}}=1  \label{1}
\end{equation}
which implies $j^{th}$ VM of $k^{th}$ user must be deployed only on one server.
\begin{gather}
C_2: \quad \sum_{j \in q}\sum_{i \in p}{V_j^{C}} \times \omega_{ji} \leq S_i^{C^{\ast}} \label{2}
\\
C_3: \quad \sum_{j \in q}\sum_{i \in p}{V_j^{M}} \times \omega_{ji} \leq S_i^{M^{\ast}} \label{3}
\end{gather}

which states that $j^{th}$ VM's CPU ($V_j^{C}$) and memory ($V_j^{M}$) requirement of  must not exceed available resource capacity of $i^{th}$ server ($S_i^{C^{\ast}}$, $S_i^{M^{\ast}}$). 

\begin{equation}
C_4: \quad \sum_{k\in m}r_k \times \tau_k \leq  \mathds{T}_k \label{5}	
\end{equation}
where, $\tau_k=t_k^{end}-t_k^{start}$ is time of execution of $k^{th}$ user's request ($r_k$). It states that $\tau_k$ must not exceed the deadline of execution i.e., $\mathds{T}_k$. 
\begin{equation}
C_5: r_k \times {R}_k \leq V_j^{R^\ast} \quad \forall_k \in [1, m], j \in [1, Q] 	\label{6}
\end{equation}
which states that required resource capacity (${R}_k $) of request $r_k$ must not exceed total available resources capacity ($R^{\ast} \in \{C^{\ast}, M^{\ast}\}$) of VM $V_j$.
\begin{equation}
C_6: \quad  \sum_{k\in m}{R}_k  \leq \sum_{i \in P}S_i^{\mathds{R^{\ast}}} \quad R^{\ast} \in \{C^{\ast}, M^{\ast}\}  \label{7}
\end{equation}
which states that aggregate of the resource capacity request of all the users must not exceed total available resources capacity of the servers altogether.
\par The considered profit maximization problem in oversubscribed industry cloud entangled with multiple constraints seeks to minimize the operational cost by reducing  the  electricity cost expenditure ($\mathds{EC}^{\mathds{CDC}}$) and improving the resource utilization ($RU^{\mathds{CDC}}$) of $\mathds{CDC}$. The terms $\mathcal{PM}$ and $r^{\mathcal{PM}}$ refer to service pricing model and service price charged for executing $r^{th}$ request, respectively. Accordingly, the resource management problem for time-slot 
$t$ is formulated as  stated in Eq. (\ref{eqmodel}):
\begin{equation}
\begin{aligned}
&	MAX: \mathds{PE} = {r^{\mathcal{PM}}\big(\omega_{kji} \big)}
\times \mathcal{PM} - {r^{\mathcal{PM}}\big(\omega_{kji} \big)}
\times {\mathds{EC}^{\mathds{CDC}}} \\
 & \text{s.t.} \quad \{C_1-C_6\} \label{eqmodel}
\end{aligned}
\end{equation}

This $\mathds{PE}$ maximization problem requires  minimization of: (\textit{i}) total power consumed (${PW^{\mathds{CDC}}}\big(\omega_{kji} \big)$) due to status of resource distribution in $\mathds{CDC}$ during time-slot $t$; (\textit{ii}) total number of active servers (${\# {\mathds{S}^{\ddagger}}}$) in $t^{th}$ time-slot accounts for energy consumption; and maximization of  resource utilization (${RU^{\mathds{CDC}}}\big(\omega_{kji} \big)$) by reducing physical resource wastage occurred due to allocation of resource capacity in excess to the actual requirement for processing user requests within $\mathds{CDC}$. The aforesaid problem  is resolved by proposing a novel OP-PMF framework described in subsequent Sections.

\section{Proposed Framework} \label{sec-proposed}
\begin{figure}[!htbp]
	\centering
	\includegraphics[width=0.9\linewidth]{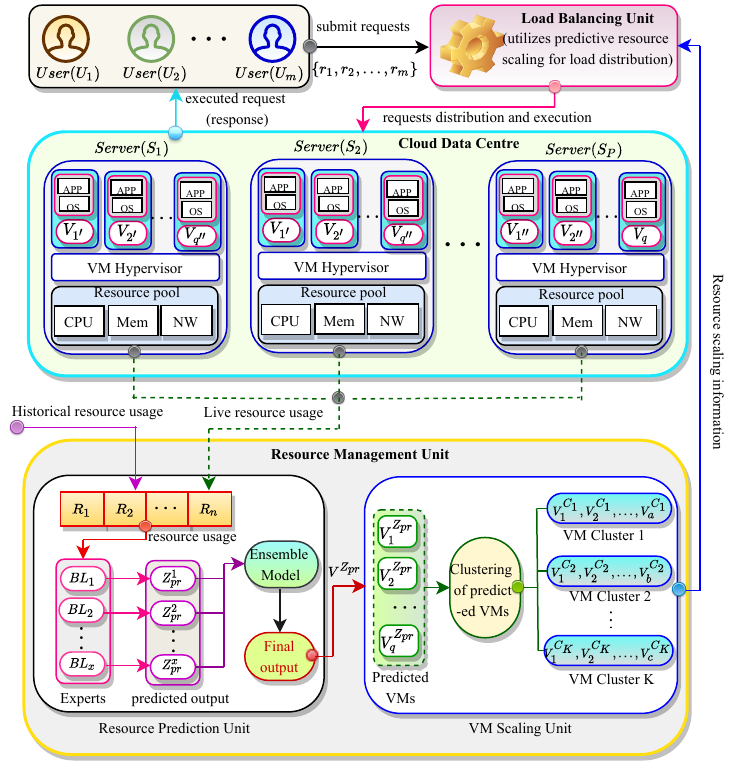}
	\caption{OP-PMF schematic overview }
	\label{fig:pm}
\end{figure}

The proposed industry cloud resource scaling framework named ``OP-PMF" comprises of \textit{Load Balancing Unit} (LBU) and \textit{Resource Management Unit} (RMU).
Let $m$ users \{${U_1, U_2, ..., U_m}$\} submit requests \{${r_1, r_2, ..., r_m}$\} to industry cloud data centres ($\mathds{CDC}$) during $t^{th}$ time-slot for execution on their purchased VMs as illustrated in Fig. \ref{fig:pm}. 
LBU is employed for distribution of users' requests \{${r_1, r_2, ..., r_m}$\} among VMs \{${V_1, V_2, ..., V_q}$\} hosted on servers \{${S_1, S_2,..., S_p}$\} in $\mathds{CDC}$. It classifies requests before execution depending on the  pricing model selected by the user (discussed in Section \ref{sec-Request Execution}). RMU operates at backend to provide knowledge of resource scaling including available resources, expected status of load, resource capacity demands, and estimated number and types of VMs to be deployed in the next time-slot $(t+1)^{th}$. Therefore, RMU and heterogeneous service model based request categorization assist LBU in making efficient resource management and load balancing decisions. RMU comprises of \textit{Resource Prediction Unit} (RPU) and \textit{VM Scaling Unit} (VSU) which operates consecutively to generate future resource scaling information. $\mathds{CDC}$ consists of clusters of servers \{${S_1, S_2, ..., S_p}$\}, where each server is comprised of VM hypervisor software layer upon which users' VMs are hosted, and pool of physical resources viz., CPU (C), Memory (M). The historical and current/live resource usage of each VM is passed to RPU for estimation of the respective VM's resource usage in the next time-slot. RPU receives combination of previous/historical and live resource usage samples i.e., total $n$ samples \{$R_1$, $R_2$, ..., $R_n$\} from $\mathds{CDC}$ for periodic training or retraining of $X$ \textit{Base Learners} (BL) \{$BL_1$, $BL_2$, ..., $BL_x$\}. The different base learners produce distinct predicted outputs ($Z^z_{pr}: \forall_z \in [1, X]$) which are used to create a vector of $X$ predicted outputs \{$Z_{pr}^1$, $Z_{pr}^2$, ..., $Z_{pr}^x$\}. Further, a computation is performed on this vector to generate an \textit{Adaptive Ensemble Model} (AEM) for production of final predicted output ($V^{Z_{pr}}$). AEM is capable of estimating accurate future resource usage of different VMs because of periodic training and re-training of $BL_1$, $BL_2$, ..., $BL_x$ and inclusion of predicted outputs from different base learners instead of depending on a single predictor's output. VSU receives a pool of predicted VMs \{$V_1^{Z_{pr}}$, $V_2^{Z_{pr}}$, ..., $V_q^{Z_{pr}}$\} from RPU.  The autoscaling of predicted VMs \{$V_1^{Z_{pr}}$, $V_2^{Z_{pr}}$, ..., $V_q^{Z_{pr}}$\} is performed by grouping them into clusters \{$\mathds{C}_1$, $\mathds{C}_2$, ..., $\mathds{C}_K$\} on the basis of respective predicted resource usage. Accordingly, $K$ clusters of VMs such as: VM Cluster 1 \{$V_1^{C_1}$, $V_2^{C_1}$, ..., $V_a^{C_1}$\}, VM Cluster 2 \{$V_1^{C_2}$, $V_2^{C_2}$, ..., $V_b^{C_2}$\}, ..., VM Cluster K \{$V_1^{C_K}$, $V_2^{C_K}$, ..., $V_a^{C_K}$\}, are determined that helps to estimate exact number and type of VMs required to serve the users' demands in the next time-slot. The detailed description of RPU, VSU and request execution are given in the subsequent sections.

\subsection{Adaptive Resource Prediction Unit} \label{sec-adaptive resource prediction}

The workflow of real-time VM resource usage prediction unit is portrayed in Fig. \ref{fig:vpu} which executes through the following steps: (\textit{i}) \textit{Data preparation}, (\textit{ii}) \textit{Base learning and measurement}, (\textit{iii}) \textit{Adaptive Ensemble Model} (AEM) generation, and (\textit{iv}) \textit{Final prediction}. The ensemble model is periodically re-created by self-adaptive weight allocation during training or re-training of real-time resource predictor.

\begin{itemize}
	\item \textit{Data prepration}: The consecutive steps of data preparation involves the \textit{aggregation} of $n$ samples of resource utilization per unit time by a VM $V_j$, \textit{rescaling} of the aggregated values in range [0, 1] using Min-Max normalization followed by an \textit{analysis} of a number of previous resource usage instances that affect the future resource usage of $V_j$. Consequently, a learning window \{$R_1^{V_j}$, $R_2^{V_j}$, ..., $R_n^{V_j}$\} composed of $n$ resource usage samples is prepared.  
	\begin{figure}[!htbp]
		\centering
		\includegraphics[width=0.85\linewidth]{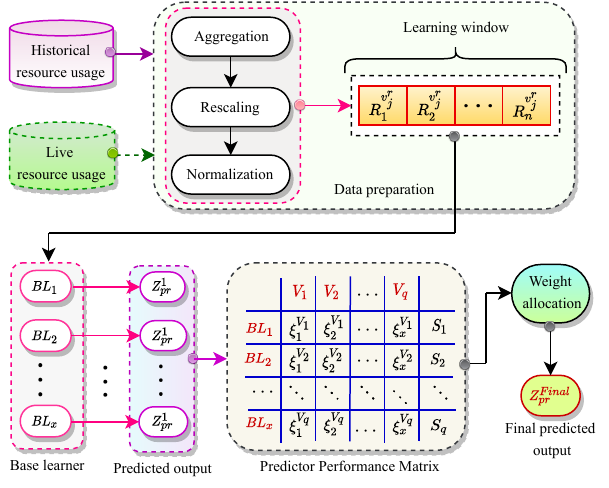}
		\caption{Adaptive ensemble learning-based  resource prediction}
		\label{fig:vpu}
	\end{figure}
	\item \textit{Base Learning}: The prepared resource usage samples \{$R_1^{V_j}$, $R_2^{V_j}$, ..., $R_n^{V_j}$\} are fed into $BL_1$, $BL_2$, ..., $BL_x$ such as Linear Regression (LR), Support Vector Machine (SVM), Neural Network (NN), Random Forest (RF) etc., to generate base prediction output in the form of a vector \{$Z_{pr}^1$, $Z_{pr}^2$, ..., $Z_{pr}^x$\}. The prediction error of each base learner is measured by computing root mean squared error ($\xi$) given in Eq. (\ref{rmse}), where $n$ is a number of data samples, $Z_{ac}$ and $Z_{pr}$ are actual and predicted outputs, respectively. 
	\begin{gather}\label{rmse}
	\xi = \sqrt{\frac{1}{n}\sum_{i=1}^{n}(Z_{ac}-Z_{pr})^2 }
	\end{gather}   
	
	\item \textit{AEM generation}: Once the initial accuracy and error measurements are produced, an ensemble model comprised of different base learners associated with different weights is generated. The mean squared error ($\xi$) is normalized in the range [0, 1] using Eq. (\ref{eqn:Normalization}), 	where $\xi_{min}$ and $\xi_{max}$ are the minimum and maximum values of the input errors, respectively. 
	\begin{equation}
	\label{eqn:Normalization}
	\hat{\xi}= \frac{ \xi_i- \xi_{min}}{\xi_{max}-\xi_{min}}
	\end{equation}
	The normalized vector $\hat{\xi}$ is a set of all normalized prediction error values for a particular resource usage. The weight score ($WS_z$) associated to $z^{th}$ base learner for $j^{th}$ VM is computed as: $WS^{V_j}_z= 1-\hat{\xi^{V_j}_z}$. A \textit{Predictor Performance Matrix} (PPM) of size ($Q \times X$) assigns weight \{$WS_1$, $WS_2$, ..., $WS_x$\} $\in \mathds{WS}$ to each base learner's prediction, where a higher value of weight indicates better performance for the current resource usage of a specific VM, and thus potentially better for the future resource usage prediction of the respective VM. The ensemble model is build using Eq. (\ref{eq:ensemble model}); where $Z^z_{pr}$
	is the prediction from $z^{th}$ base learner. 
	
	\begin{equation}
	\textit{AEM}: \frac{\sum_{z=1}^{X}{(WS^{V_j}_z \times Z^z_{pr})}}{\sum_{z=1}^{X}{WS^{V_j}_z}}
	\label{eq:ensemble model}
	\end{equation}

	\item \textit{Final prediction}: The final predicted output is potentially more accurate because this ensemble model assigns higher weights and lower weights to potentially more accurate
	and less accurate base learners, respectively. Also, the predictions of base learners in the ensemble model are updated to historical data for future evaluation of the base learners. 
	
\end{itemize}

Algorithm \ref{algo-prediction} depicts summarized steps of real-time VM Prediction. Step 1 receives list of active users' VMs for resource usage prediction that requires complexity of $O(1)$. The steps 2-9 iterate for $Q$ (i.e., total number of VMs) times, wherein steps 3-8 repeat for $x$ (i.e., number of base learners) intervals. Since, each base learner has different time-complexity,  $\mathds{Z}$ is a time-complexity variable for base learner ($BL_z$). Steps 10-12 produce final predicted output by generating \textit{Adaptive Ensemble Model} using Eq. (\ref{eq:ensemble model}) which repeat $Q$ times with complexity $O(Qx)$. The overall complexity of Algorithm \ref{algo-prediction} is $O(Q^2x^2\mathds{Z})$.     
\begin{figure}[!htbp]
	\removelatexerror
	\begin{algorithm}[H]
		\caption{Ensemble approach for VM Prediction}
		\label{algo-prediction}
		\SetKwFunction{FMain}{VM Prediction}                                                                                
		\SetKwProg{Fn}{Function}{:}{}  
		\Fn{\FMain{$List_{\mathds{V}}$ }} {                                       \For{each VM $V_j$}{                                                                      	\For{each Base learner $BL_z$}{
					$V_j^{Z_{pr}}$=$BL_z$($V_j^{C}$, $V_j^{M}$)\;  
					$List_{\mathds{V}^{Z_{pr}}}$  $\Leftarrow$  $V_j^{Z_{pr}}$\;                                                          			Compute error ($\xi_{jz}$) using Eq. (\ref{rmse})\;
					Update Predictor Performance Matrix: $PPM[j][z]= 1-\xi_{jz}$\;
			} } 
			\For{each VM $V_j$}{
				$List_{\mathds{V}^\ast}$ $\Leftarrow$	\textit{Ensemble Model($PPM$, $List_{\mathds{V}^{Z_{pr}}}$)}\;
			}                                                                                                                                                                                                          		
			\Return{ $List_{\mathds{V}^\ast}$}                                                                         }

	\end{algorithm}
	
\end{figure}

\subsection{Leveraging Cloud Oversubscription } \label{sec-leveraging}
\begin{figure}[!htbp]
	\centering
	\includegraphics[width=0.85\linewidth]{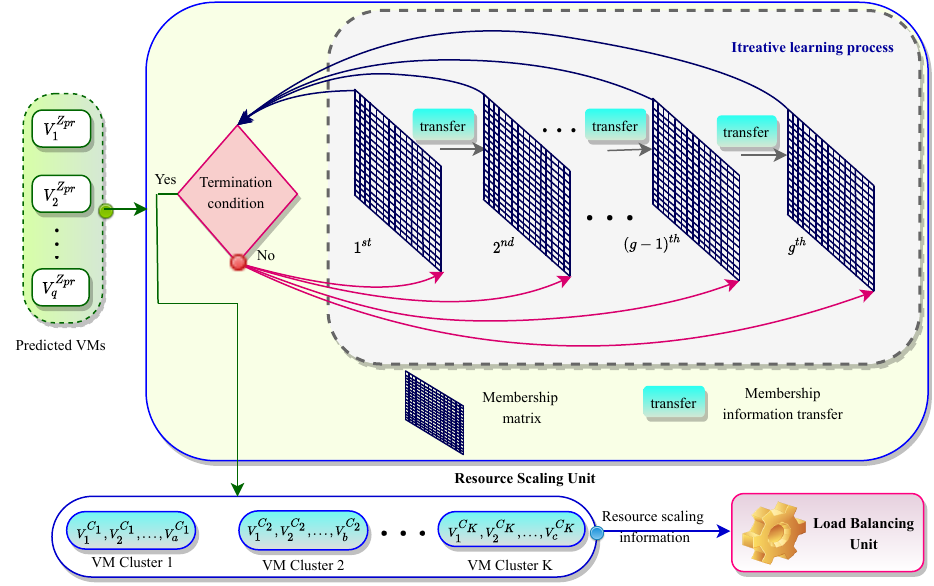}
	\caption{VM Scaling Unit}
	\label{fig:rsu}
\end{figure}

The cloud oversubscription is leveraged by allocating the physical resources to the VMs respective to the predicted VMs instead of demanded resource capacities. Further, the predicted VMs of similar resource requirement are grouped using Fuzzy C-means clustering to determine the exact number and size/type of VMs required to execute future workload in the next time-slot. Let $\mathds{V}^\ast =$ \{$V_1^{Z_{pr}}$, $V_2^{Z_{pr}}$, ..., $V_q^{Z_{pr}}$\} be the set of predicted VMs and $\mathds{C}^\ast=$ \{$\mathds{C}^\ast_1$, $\mathds{C}^\ast_2$, ..., $\mathds{C}^\ast_K$\} is a set of centers of VM clusters \{$\mathds{C}_1$, $\mathds{C}_2$, ..., $\mathds{C}_K$\} as shown in Fig. \ref{fig:rsu}. 
A \textit{membership value} ($\mu$) is assigned to each predicted VM by computing euclidean distance ($\delta_{jy}$) between cluster's center ($\mathds{C}^\ast_y$) and predicted VM ($V_j^{Z_{pr}}$). 
The lesser distance ($\delta_{jy}$) indicates higher membership of $V_j^{Z_{pr}}$ towards $\mathds{C}^\ast_y$ i.e., predicted VM is more closer to the particular cluster center. During iterative learning process, membership ($\mu_{jd}$) of each predicted VM ($V_j^{Z_{pr}}$) within the cluster ($\mathds{C}_d$) and cluster center ($\mathds{C}^\ast_d$) are updated using Eqs. (\ref{eq:membership}) and (\ref{eq:clustercenter});
\begin{equation}
\mu_{jd}= \frac{1}{\sum_{y=1}^{K}{(\frac{\delta_{jd}}{\delta_{jy}})}^{(2/\mathds{M}^\ast -1)}} \label{eq:membership}	
\end{equation}

\begin{equation}
\label{eq:clustercenter}
\mathds{C}^\ast_d = \frac{\sum_{j=1}^{Q}{((\mu_{jd})^\mathds{M} \times V_j^{Z_{pr}})}}{\sum_{j=1}^{Q}{(\mu_{jd})^\mathds{M}}}
\end{equation}
where $\forall_{d} \in \{1, 2, ..., K\}$, $\forall_{j} \in \{1, 2, ..., Q\}$, $\mathds{M} \in [1, \infty]$ is fuzziness index. The objective function given in Eq. (\ref{fuzzy}) iterates to make the inter-cluster resource usage similar, while keeping the clusters as different (far) as possible.
\begin{equation}
\label{fuzzy}
Minimize \sum_{j=1}^{Q}\sum_{d=1}^{K}(\mu_{jd})^\mathds{M} ||V_j^{Z_{pr}} - \mathds{C}^\ast_d||^2
\end{equation}
where $||V_j^{Z_{pr}} - \mathds{C}^\ast_d||^2$ is euclidean distance between predicted VM ($V_j^{Z_{pr}}$) and cluster center ($\mathds{C}^\ast_d$). The size of VM means processor and memory capacity of VM that determines the physical configuration of the respective VM. All the predicted VMs of a cluster are assigned resource capacity by comparing the maximum and minimum size of the predicted VMs in the specific cluster with the available size of VMs' configuration. Accordingly, the exact number of a particular size of VMs required to execute users' request in future, is determined by comparing clusters \{$\mathds{C}_1$, $\mathds{C}_2$, ..., $\mathds{C}_K$\} of predicted VMs to a specific VM type having fixed resource \{$C$, $M$\} capacity using Eq. (\ref{eq.autoscale}); 

\begin{equation}\label{eq.autoscale}
V^{type}_{\mathds{C}_y} =
\begin{cases}
V_{\mathds{SM}}, & {(\mathds{C}^{R_{MAX}}_{y} \leq V_{\mathds{SM}}^{{R}} )} \\
V_{\mathds{ME}} , & {( V_{\mathds{SM}}^{{R}} < \mathds{C}^{R_{MIN}}_{y} \wedge \mathds{C}^{R_{MAX}}_{y} \leq V_{\mathds{ME}}^{{R}})} \\
V_{\mathds{LA}} , & {(V_{\mathds{LA}}^{{R}} <\mathds{C}^{R_{MIN}}_{y} \wedge \mathds{C}^{R_{MAX}}_{y} \leq V_{\mathds{LA}}^{{R}})} \\
V_{\mathds{XL}} , & {(otherwise)} 
\end{cases} 
\end{equation}
where $\quad {R} \in \{C, M\}$, $ V_{\mathds{SM}}^{{R}}$, $ V_{\mathds{ME}}^{{R}}$, $V_{\mathds{LA}}^{{R}}$ and $V_{\mathds{XL}}^{{R}}$ represents small, medium, large and extra-large types of VMs, respectively having resource capacity as per their respective type, and $\mathds{C}^{R_{MAX}}_{y}$ and $\mathds{C}^{R_{MIN}}_{y}$ are maximum and minimum resource utilization of cluster $\mathds{C}_y$. If the maximum resource requirement of a VM from $\mathds{C}_y$ is lesser or equals to the resource capacity of $V_{\mathds{SM}}$, then small type of VM is mapped to $\mathds{C}_y$ and selected for the future requests execution. The required number of VMs is equal to the total number of predicted VMs in the respective cluster.

\subsection{Exploiting Service Pricing Models} \label{sec-Request Execution}
\begin{figure}[!htbp]
	\centering
	\includegraphics[width=0.9\linewidth]{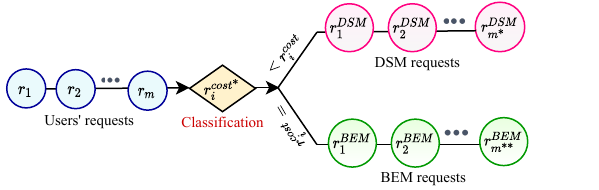}
	\caption{Request Categorization}
	\label{fig:requestexecution}
\end{figure}

The cloud users' request \{$r_1$, $r_2$, ..., $r_m$\} is categorized into two divisions according to the service pricing model selected by the cloud user ($i$) \textit{Delay Sensitive Model} (DSM) and ($ii$) \textit{Best-Effort Model} (BEM) as shown in Fig. \ref{fig:requestexecution}. DSM specifies the request that must be executed within deadline and the cloud user agree to pay more for their completion before deadline.
BEM states that the request must be completed within deadline and the charges are fixed depending on resource requirement of request for execution. Eq. (\ref{req}) determines the class of request $r^{class}_k$, where $\tau_k=t_k^{end}-t_k^{start}$ is the  execution time of $r_k$ and $ U_k^{\ddagger}$ states that user $U_k$ is agree to pay more for faster execution of request $r_k$.

\begin{equation}
r_k^{class} = \begin{cases}
r_k^{DSM}, & {If(r_k \times \tau_k <\mathds{T}^{\ast}_{r_k} \wedge U_k^{\ddagger})} \\
r_k^{BEM}, & {\text{otherwise}}  
\end{cases}	
\label{req}	
\end{equation}

During time-slot $t$, all the VMs executing requests subject to DSM \{$r^{DSM}_1$, $r^{DSM}_2$, ..., $r^{DSM }_{m^\ast}$\} $\in {r}^{DSM}$, are hosted on servers having high processing speed to maximize the profit. The requests belonging to BEM \{$r^{BEM}_1$, $r^{BEM}_2$, ..., $r^{BEM}_{m^{\ast\ast}}$\} $\in {r}^{BEM}$ are preferably hosted on low speed server (with lesser execution cost) to minimize power consumption and operational cost.
A mapping $\omega|\omega_{ji}: U_k \times V^{r_{k}}_j \mapsto S_i$ defines allocation of $j^{th}$ VM executing request ($r_{k}$) of user ($U_k$) to server $S_i$. A $j^{th}$ VM ($V_j$) is deployed on server ($S_i$) if it satisfies the constraints stated in Eqs. (\ref{2}) and (\ref{3}), which states that $j^{th}$ VM's CPU ($V_j^{C}$) and memory ($V_j^{M}$) requirement   must not exceed the available respective resource capacity of $i^{th}$ server ($S_i^{C^{\ast}}$, $S_i^{M^{\ast}}$).
\begin{equation}
 \sum_{j \in Q}{V_j^{C}} \times \omega_{ji} \leq S_i^{C^{\ast}}  \quad \forall_{i \in \{1, 2, ..., P\}}\label{2}
\end{equation}

\begin{equation}
\sum_{j \in Q}{V_j^{M}} \times \omega_{ji} \leq S_i^{M^{\ast}}\quad \forall_{i \in \{1, 2, ..., P\}} \label{3}
\end{equation}
 For a time-slot \{$t_1$, $t_2$\}, the user requests are executed as follows:
\begin{itemize}
	\item All the DSM based requests \{$r^{DSM}_1$, $r^{DSM}_2$, ..., $r^{DSM \ast}_m$\} are scheduled on VMs for execution by sorting them in increasing order of their deadline in the list of VMs ($List_{\mathds{V}^{Dealine^{+}}}$), where the mapping $r^{DSM}_k \times V_j$ is an element in the list $List_{\mathds{V}^{Dealine^{+}}}$ which represents that $r^{DSM}_k$ is hosted on VM $V_j$ for execution. All the available servers are sorted in decreasing order of their processing speed in the list ($List_{\mathds{S}^{C^{-}}}$). The VMs executing requests in the increasing order from $List_{\mathds{V}^{Dealine^{+}}}$ are assigned to server in first-fit decreasing order of their processing speed in the sorted list $List_{\mathds{S}^{C^{-}}}$. Eq. (\ref{eq:dsm}) states allocation of $r^{DSM}_k \times V_j$ to active server having maximum processing speed from $List_{\mathds{S}^{C^{-}}}$. Such an allocation of VMs executes lowest deadline requests on highest processing speed servers to minimize time of execution and maximize profit for the $\mathds{CDC}$ by increasing the pricing charges for user. 
	\begin{equation}
	\omega_{ji}^{r^{DSM}_k}=r^{DSM}_k \times V_j \times Max(List_{\mathds{S}^{C^{-}}}) \label{eq:dsm}
	\end{equation}
	
	\item The remaining active servers are arranged in increasing order of 
	their processing speed in the sorted list $List_{\mathds{S}^{C^{+}}}$ and all the VMs deploying BEM based requests \{$V^{r^{BEM}_1}$, $V^{r^{BEM}_2}$, ..., $V^{r^{BEM }_{m^{\ast\ast}}}$\} are arranged in decreasing order of their processing requirement in the sorted list $List_{\mathds{V}^{C^{-}}}$. The VMs executing high computing requests from $List_{\mathds{V}^{C^{-}}}$ are allocated to slow speed servers in the list ($List_{\mathds{S}^{C^{+}}}$) so as to minimize the cost of execution as stated in Eq. (\ref{eq:bem}). Consequently, VMs executing requests with high resource requirement on low processing speed servers would minimize power consumption (as slow speed servers consumes less power as compared to high speed servers of same resource capacity) and electricity cost for $\mathds{CDC}$. 
	\begin{equation}
	\omega_{ji}^{r^{BEM}_k}=r^{BEM}_k \times V_j \times Min(List_{\mathds{S}^{C^{+}}})  \label{eq:bem}
	\end{equation}
	
\end{itemize}

\section{Operational Design and Illustration} \label{sec-operational design}
\subsection{Algorithm and Complexity}
Algorithm \ref{algo-OP-PMF} outlines the operational procedure of OP-PMF. It initiates sets of users \{$U_1$, $U_2$, ..., $U_m$\}, servers \{$S_1$, $S_2$, ..., $S_p$\}, and users' VMs \{$V_1$, $V_2$, ..., $V_q$\}. The historical CPU and memory utilization of each user's VM \{$V_j^{{R}_1}$, $V_j^{{R}_2}$, ..., $V_j^{{R}_n}$\}$: \forall_{j}\in \{1, 2, ..., q\} $ are entered into the VM prediction unit for proactive estimation of their CPU ($V_j^{{Z}{pr}^{C}}$) and memory ($V_j^{{Z}{pr}^{M}}$) utilization. Subsequently, predicted VMs are clustered into $K$ groups based on their estimated CPU and memory usage to ascertain the sizes and numbers of VMs needed for the subsequent time-slot. During consecutive time-slots \{$t_1$, $t_2$\}, distinct requests \{$r_1$, $r_2$, ..., $r_m$\}  with deadlines and a random tag $DSM$ or $BEM$ are generated for each user. These requests are then allocated to respective users' VMs for execution, adhering to the approach delineated in Section \ref{sec-Request Execution}.
\begin{figure}[!htbp]
	\removelatexerror
	\begin{algorithm}[H]
		\caption{OP-PMF Operational Summary}
		\label{algo-OP-PMF}
		Initialize: $List_{{\mathds{S}}}$, $List_{\mathds{V}}$, $List_{\mathds{U}}$\; 
		
		\For {each time-interval $\{t_1, t_2\}$}{ 
			
			{$List_{\mathds{V}^\ast}$ $\Leftarrow$ VM Prediction($List_{\mathds{V}}$)\;
				
			}
			Group predicted VMs into clusters: \{$\mathds{C}_1$, $\mathds{C}_2$, ..., $\mathds{C}_K$\} $\Leftarrow$ VM Clustering($List_{\mathds{V}^\ast}$)\;
			\For{each $d^{th}$ cluster:}
			{$V_j^{C}$ = Max($\mathds{C}^C_y$)\;
				$V_j^{M}$ = Max($\mathds{C}^M_y$)\;
				Selected size of VM ($V^{Type}$)$\Leftarrow$ ($V_j^{C}$, $V_j^{M}$)\;
				Required number of $V^{Type}$ = Size($\mathds{C}_y$)\; 
			}	
			
		}
		
		\For {each time-interval $\{{t}_1+1, {t}_2+1\}$}{
			Prepare list of request based on DSM: $List_{r^{DSM}: \{{R}^{DSM},t^{start}, t^{end}\}}$ and $List_{r^{BEM}: \{{R}^{BEM},t^{start}, t^{end}\}}$\;
			Sort all the available server ($List_{{\mathds{S}}}$) in decreasing order of processing speed\;
			Assign VMs loaded with request ${r^{DSM}}$ on high-speed processing server that satisfy all the constraints (Eqs. (\ref{2}) and (\ref{3}))\;
			Sort list of all remaining active servers ($List_{{\mathds{S}}}$) in increasing order of processing speed\;
			Assign VMs loaded with request ${r^{BEM}}$ on low-speed processing server that satisfy all the constraints (Eqs. (\ref{2}) and (\ref{3}))\;
		}

	\end{algorithm}
	
\end{figure}
Step 1 initializes lists of servers, users, and VMs associated to different users consumes time-complexity $O(1)$. Let steps 2-11 repeat for $t$ time-slots, where step 3 calls Algorithm \ref{algo-prediction} i.e., ensemble approach based VM prediction unit consumes time-complexity of $O(Q^2x^2\mathds{Z})$. Step 4 arranges predicted VMs into clusters by applying concepts of \textit{Fuzzy C-means Clustering} shows complexity $O(\mathds{C})=O(NK\mathds{I})$, where $N$ is the number of links, $K$ is the number of clusters, and $\mathds{I}$ is the number of iterations to run by the procedure. Steps 5-10 iterate for $K$ (i.e., number of clusters) intervals to estimate the size and number of different type of VMs ($V^{Type}$) to be required in next time-slot, consumes time $O(K)$. Further, steps 12-18 repeat for $t+1$ intervals, wherein, step 14 has time-complexity of $O(P^2)$ ($P$ is total number of servers) while step 15 consumes time equivalent to $O(P\times Q^{DSM})\simeq O(P\times Q) $ for allocation of all VMs hosting $DSM$ based requests. Similarly, steps 16 and 17 consume complexity of $O(P_{rest}^2)$ ($P_{rest}$ is total number of remaining active servers) and $O(P_{rest}\times Q^{BEM}) \simeq O(P\times Q)$, respectively. The complete time-complexity of OP-PMF approach is $O(P^2Q^2x^2\mathds{Z}\mathds{C})$.

\subsection{Numerical Illustration}
Consider two industries A and B operates using Best Effort model and Delay sensitive model, respectively, from industry cloud provider (ICP). Let industry A hired four VMs (CPU in MIPS, memory in GB): $V_1^{A}$(500, 0.5), $V^{A}_2$(500, 0.5),  $V^{A}_3$(1000, 1), $V^{A}_4$(2000, 3); and   industry B hired three VMs (CPU in MIPS, memory in GB): $V^{B}_1$(500, 0.5), $V^{B}_2$(1000, 0.5),  $V^{B}_3$(1500, 2). There previous VM usage are given in Table \ref{table:illus}. 
\begin{table}[!htbp]
	\centering
	
	\caption[Table caption text] {VM CPU (MIPS), memory (GB) Usage}  
	\label{table:illus}
	\resizebox{9.1cm}{!}{
		\begin{tabular}{|lcc|}
			\hline
			Time&Industry A& Industry B\\
			\hline
		$t_1$&	$V^{A}_1$(200, 0.1), $V^{A}_2$(300, 0.3),  $V^{A}_3$(400, 0.3), $V^{A}_4$(800, 1) & $V^{B}_1$(300, 0.2), $V^{B}_2$(420, 0.3),  $V^{B}_3$(1000, 0.8) \\
			
		$t_2$&	$V^{A}_1$(300, 0.3), $V^{A}_2$(200, 0.2),  $V^{A}_3$(450, 0.4), $V^{A}_4$(500, 0.3) & $V^{B}_1$(200, 0.4), $V^{B}_2$(800, 0.3),  $V^{B}_3$(700, 0.6) \\
		$\ldots$ & $\ldots$ &$\ldots$ \\
		$t_m$&	$V^{A}_1$(400, 0.2), $V^{A}_2$(100, 0.2),  $V^{A}_3$(600, 0.5), $V^{A}_4$(800, 1) & $V^{B}_1$(250, 0.2), $V^{B}_2$(300, 0.2),  $V^{B}_3$(800, 1) \\	
	
		$\ldots$ & $\ldots$ &$\ldots$ \\
		$t_n$&	$V^{A}_1$(300, 0.2), $V^{A}_2$(250, 0.2),  $V^{A}_3$(400, 0.4), $V^{A}_4$(1000, 0.8) & $V^{B}_1$(300, 0.2), $V^{B}_2$(700, 0.3),  $V^{B}_3$(900, 0.9) \\	
			\hline
	\end{tabular}}
\end{table}
The VM configuration are reffered from Table \ref{table:vm} while server configuration and power consumption are applied from Table \ref{table:server} during allocation of these VMs. The electricity cost expenditure ($\mathds{EC}$) is estimated using $\mathds{CDC}$ electricity price values equals to $0.07 \$/KWH$, which is obtained on the basis of average electricity price in U.S. \cite{usEnergy2014}. Referring Delay sensitive model, the extra charge for earlier execution is 0.25\$/H.
\textit{Deploying OP-PMF approach, how much power consumption and electricity bill of ICP is reduced}?  \textit{Also, compute the quantitative profit maximization achieved by ICP}? 

{\textit{Solution}:} For the given scenario, applying OP-PMF, these seven VMs have predicted resource usage as $V^{A}_1$(280, 0.1), $V^{A}_2$(290, 0.3),  $V^{A}_3$(300, 0.3), $V^{A}_4$(930, 0.7),  $V^{B}_1$(380, 0.2), $V^{B}_2$(700, 0.6),  $V^{B}_3$(930, 0.4). The average resource utilization  rate of the servers is 35\%-50\% only. Accordingly, they are kept in cluster such as $\mathds{C}_1:$ \{$V^{A}_1$(280, 0.1), $V^{A}_2$(290, 0.3),  $V^{A}_3$(300, 0.3), $V^{B}_1$(380, 0.2)\} $\in V_{\mathds{SM}}$ and $\mathds{C}_2:$ \{$V^{A}_4$(930, 0.7),  $V^{B}_2$(700, 0.6),  $V^{B}_3$(930, 0.4)\} $\in V_{\mathds{ME}}$. These 7 VMs are hosted on one server $S_1$ (Table \ref{table:server}) with 2 processing elements (PE).
\\ Applying Eqs. (\ref{RU1}), (\ref{RU2}) and (\ref{power2}), the power consumption is equals to 2158.7 KW with resource utilization of 50\%. The electricity cost expenditure is approx. 151.109 \$/H. 
\\Otherhand, applying classical VM placement approach, there are 3 VMs $\in V_{\mathds{SM}}$, 2 VMs $\in V_{\mathds{ME}}$, 1 VM $\in V_{\mathds{LA}}$, and $\in V_{\mathds{XL}}$. These VMs can be accommodated either on 2 $S_1$ or 1 $S_2$ servers, where the power consumption will be equals to 4317.4 KW or 3539.23 KW. The electricity bill will cost approx. 302.208\$/H or 247.746 \$/H. 
Therefore, using OP-PMF, the electricity bill is reduced up to 49.4\% for ICP. Since industry A operates 4 VMs using BEM and industry B operates 3 VMs using DSM, there are 42.2\% VMs operating with the privilege of earlier execution (let say, 10 minutes) which results into the earned profit is 1.76\$ per hour. The achieved performance and profit gain for ICP will be much higher in the real scenario with multiple number of VM demands with varying service pricing models from thousands of industries.

\section{Performance Evaluation and Comparison} \label{sec-performance}

\subsection{Experimental Set-up}
The simulation experiments are executed on a server machine assembled with two Intel\textsuperscript{\textregistered} Xeon\textsuperscript{\textregistered} Silver 4114 CPU with 40 core processor and 2.20 GHz clock speed. The computation machine is deployed with 64-bit Ubuntu 16.04 LTS, having main memory of 128 GB. The data center environment was set up with three different types of server and four types of VMs configuration shown in Tables \ref{table:server} and \ref{table:vm} in Python version-3. The resource features like power consumption ($P_{max}, P_{min}$), MIPS, RAM and memory are taken from real server IBM \cite{IBM1999} and Dell \cite{Dell1999} configuration where $S_1$ is `ProLiantM110G5XEON3075', $S_2$ is `IBMX3250Xeonx3480' and $S_3$ is `IBM3550Xeonx5675'. Furthermore, the experimental VMs configuration are inspired from the VM instances of Amazon website \cite{amazon1999EC2}. 

\begin{table}[!htbp]
	\centering
	
	\caption[Table caption text] {Server Configuration}  
	\label{table:server}
	\resizebox{9.2cm}{!}{
		\begin{tabular}{|lcccccc|}
			\hline
			Server&PE&MIPS&RAM(GB)&Storage (GB)&$PW_{max}$&$PW_{min}$/$PW_{idle}$\\
			\hline
			$S_1$ 	& 2&2660&4&160&135&93.7 \\
			$S_2$	& 4&3067&8&250&113&42.3 \\
			$S_3$	& 12&3067&16&500&222&58.4 \\

			\hline
	\end{tabular}}
\end{table}

\begin{table}[htbp]
	\centering
	
	\caption[Table caption text] {VM Configuration}  
	\label{table:vm}
	\begin{tabular}{|lcccc|}
		\hline
		VM type& PE &MIPS&RAM(GB)&Storage (GB)\\
		\hline
		$V_{\mathds{SM}}$&1&500&0.5&40\\
		$V_{\mathds{ME}}$&2&1000&1&60\\
		$V_{\mathds{LA}}$&3&1500&2&80\\
		$V_{\mathds{XL}}$&4&2000&3&100\\

		\hline
	\end{tabular}
\end{table}


\textit{{Datasets}}: The resource utilization for different VMs followed from two real workloads including Google Cluster Data (GCD) \cite{reiss2011google} and PlanetLab VMs traces (PL) \cite{beloglazov2012optimal}. GCD has resources CPU, memory, disk I/O  request and usage information of 672,300 jobs executed on 12,500 servers for the period of 29 days. The CPU and memory utilization percentage of VMs are obtained from the given CPU and memory usage percentage for each task in every five minutes. PL contains CPU utilization of more than 11K VMs measured every five minutes during ten random days in March-April, 2011. We extracted CPU and memory usage percentages from GCD and CPU percentage from PL as per their availability for various experiments.

The original GCD and PL traces of VMs only reports resource utilization percentage of VMs per unit time without any information about number of users, deadline and charges of request execution. Therefore, we generated users equals to 60\% of the total number of VMs and mapped random number of VMs in the range [1-5] to them such that each user owns atleast one VM and all the VMs are owned by some user. To decide the heterogeneous requests ($r^{DSM}$ or $r^{BEM}$) from user, a random generator is used to produce different number (such as 20\%, 40\%, 60\%, 80\%, 100\%) of $r^{DSM}$ and the remaining requests are taken as $r^{BEM}$. The time-interval of real-time VM prediction and processing is 5 minutes which is set according to the time-interval of resource usage records available from both datasets.  

\textit{Comparative Approaches}: 
The proposed work is compared for different performance metrics with various state-of-art approaches including Slack and Battery Aware placement (SBA) \cite{alanazi2017reducing}, Static THReshold with Multiple Usage Prediction (THR-P) and Dynamic threshold based on Local Regression with Multiple Usage Prediction (LR-P) \cite{hieu2017virtual}, Online VM Prediction based Multi-objective Load Balancing (OP-MLB) \cite{saxena2021op}, First-Fit \cite{jangiti2019aggregated}, Random-Fit \cite{jung2010mistral} and Best-Fit \cite{shirvastava2017best} heuristics. Furthermore, overall electricity cost and profit earned using OP-PMF  with over-subscription (OP-PMF$^+$) compared to without OP-PMF  (i.e., OP-PMF$^-$) is also presented. Table \ref{KPI} presents comparison of key performance indicators of proposed framework versus comparative approaches.

\begin{table}[!htbp]
	
	\caption{{Key Performance Indicators Analysis }}
	\label{KPI}
	\scriptsize
	\centering
	\resizebox{0.5\textwidth}{!}{
		\begin{tabular}{|l|c|c|c|c|c|c|c|c|c|}
			\hline
			{	\textbf{KPI}}&{\textbf{\cite{saxena2021op}}}&{\textbf{\cite{dabbagh2016energy}}}&{\textbf{\cite{dabbagh2015exploiting}}}&{\textbf{\cite{alanazi2017reducing}}}&\textbf{\cite{sahoo2017lvrm}}& {\textbf{\cite{hieu2017virtual}}}& \textbf{\cite{jangiti2019aggregated}}&\textbf{\cite{shirvastava2017best}}& \small{\textbf{OP-PMF}}\\ \hline \hline
			
			$\mathds{V}^{Pr}$&{\checkmark } &{$\checkmark$}  &{$\times$}&{$\times$}&{$\times$} &{$\times$} & $\times$ &$\times$ &\checkmark \\ \hline
			$\mathds{V}^{Sc}$& {$\times$}&{$\times$} &{$\times$}&{$\times$}&{$\times$}&{$\times$}&$\times$ &$\times$ &\checkmark \\ \hline
			$\mathds{EC}$	& {$\times$}&$\times$ &{\checkmark}&{\checkmark}&{$\times$} &{$\times$}& $\times$ &$\times$ &\checkmark \\ \hline
			$r^{class}$&{$\times$}&$\times$ &{\checkmark }&{$\times$}&{$\times$} &{$\times$}&$\times$ & $\times$&  \checkmark\\ \hline
			
			$\mathds{O}_{\mathds{S}}$ &{\checkmark} &{\checkmark}  &{$\times$}&{$\times$}&{$\times$}& $\checkmark$ &$\times$ & $\times$&\checkmark\\ \hline
			$\#\mathds{S}^{\ddagger}$ & {\checkmark} &{\checkmark}  &{\checkmark}&{\checkmark}&{\checkmark } &{\checkmark} &\checkmark & $\times$& \checkmark\\ \hline
			$\mathds{PR}$ &{$\times$} &{ $\times$} &{\checkmark}&{$\times$}&{$\times$} &{$\times$}  &$\times$ & $\times$& \checkmark\\ \hline
			$RU$ &{\checkmark} &{\checkmark } &{\checkmark}&{$\checkmark$}&{$\checkmark$} &{$\times$}  &\checkmark &$\times$ &\checkmark\\ \hline
			$PW$ &{\checkmark} &{\checkmark } &{\checkmark}&{$\checkmark$}&{$\checkmark$} &{\checkmark}  & \checkmark&\checkmark & \checkmark \\ \hline
		\end{tabular}
	}
	\\	\footnotesize{\tiny{{$\mathds{V}^{Pr}$: VM Prediction , $\mathds{V}^{Sc}$: VM Autoscaling, $\mathds{EC}$: Electricity Cost, $r^{class}$: Requests Classification,\\ $\mathds{O}_{\mathds{S}}$: Overloads, $\#\mathds{S}^{\ddagger}$: Active servers, $\mathds{PR}$: Earned Profit, $RU$: Resource utilization, $PW$: Power consumption}}}
\end{table} 
\subsection{Resource Prediction} 

\subsubsection{Accuracy}
\begin{figure}[!htbp]
	\centering
	\includegraphics[width=0.65\linewidth]{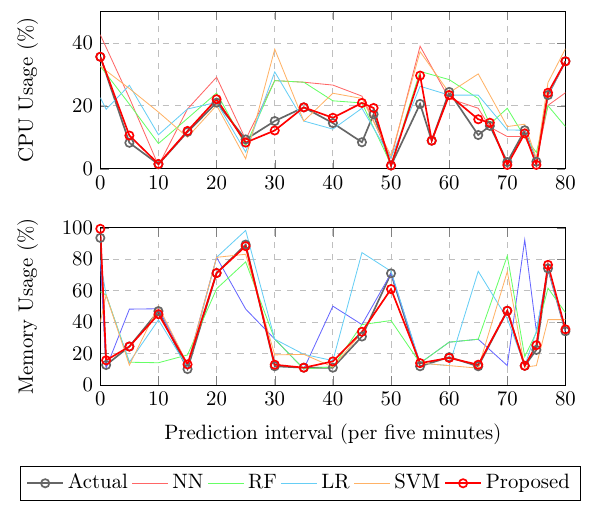}
	\caption{Google Cluster: VM Resource Prediction}
	\label{fig:gcdcpumemory}
\end{figure}

\begin{figure}[!htbp]
	\centering
	\includegraphics[width=0.65\linewidth]{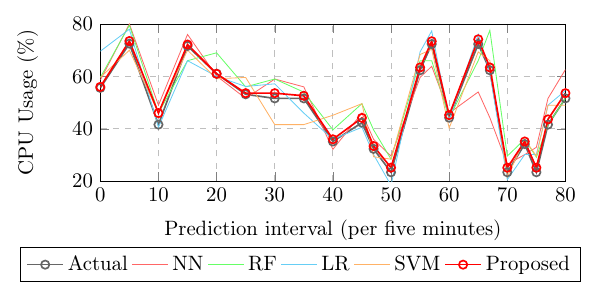}
	\caption{Planet Lab: VM Resource Prediction}
	\label{fig:plcpu}
\end{figure}

\begin{figure*}[!htbp]
	\centering	
	\subfigure[Google Cluster-CPU]{\includegraphics[width=0.31\linewidth, scale=2]{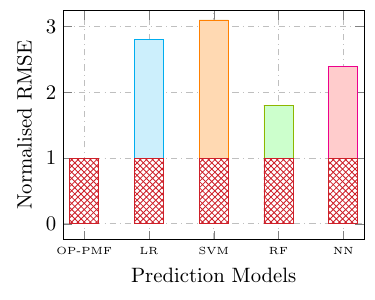}} 
	\subfigure[Google Cluster-Memory]{\includegraphics[width=0.31\linewidth, scale=2]{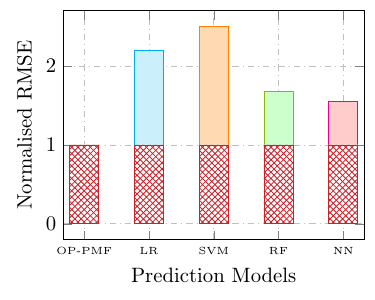}} 
	\subfigure[Planet Lab-CPU]{\includegraphics[width=0.31\linewidth, scale=2]{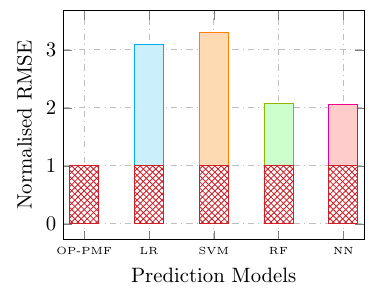}} 	
	\caption{Normalised RMSE}
	\label{Normalised RMSE}
	
\end{figure*}

The performance evaluation of OP-PMF initiates with investigation of RPU. Figs. \ref{fig:gcdcpumemory} and \ref{fig:plcpu} compare actual versus predicted accuracy of VM resource usage traces of GCD and PL workloads, respectively. During simulation, the predicted resources usage of each VM from both dataset is recorded and utilized to evaluate efficiency of OP-PMF . However, the CPU and memory usage of a randomly selected VM from GCD and PL are shown in Figs. \ref{fig:gcdcpumemory} and  \ref{fig:plcpu}. It is to be observed that predicted CPU and memory usage have almost overlapped the actual CPU and memory usage for VMs of both the traces in case of proposed approach. While the other comparative predictors including neural network (NN), Random Forest (RF), Linear Regression (LR), and Support Vector Machine (SVM) show fluctuating performance over the time-period.  
The normalised RMSE of different prediction models is compared with respect to Average RMSE values of proposed approach in Fig. \ref{Normalised RMSE}, where the normalised value 1 means best and greater values depict worse performance.
The reason behind high and consistent accuracy of the proposed prediction approach is that it do not depends on prediction of a single predictor model, instead it captures best predicted values resulted from analyses of predicted output of different base predictors. It learns live and historical resource usage patterns, correlations among them, and retrains itself according to the changes in respective resource usage of each VM periodically which is responsible for near accurate prediction of resources.  
\subsubsection{Number of Overloads}
\begin{figure}[!htbp]
	
	\centering
	\subfigure[Google Cluster ]{\includegraphics[width=.23\textwidth]{ 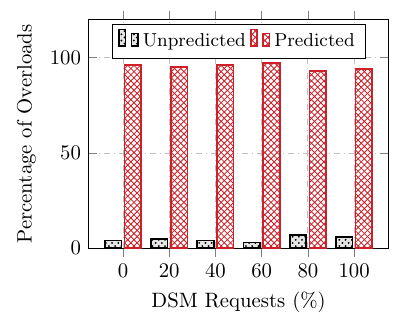}}
	\subfigure[Planet Lab ]{\includegraphics[width=.23\textwidth]{ 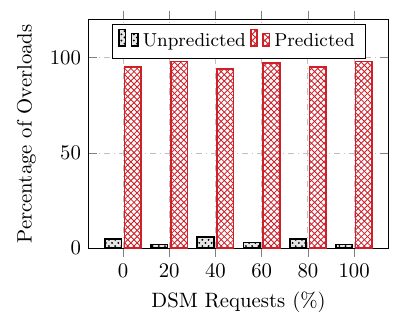}}\hfill
	
	\caption{Overloads}
	\label{overloads}
	
\end{figure}

The number of unpredicted overloads prompt unavailability of servers leading to SLA violation and VM migration that accounts for further power consumption and higher electricity cost. Fig. \ref{overloads} shows the predicted and unpredicted overloads for both VM traces. It is observed that percentage of correctly predicted overloads is influentially high (i.e., up to 99.6\%). However, the number of unpredicted overloads are either lesser or equal to 2.3\% which is independent of the ratios of $r^{DSM}:r^{BEM}$ for each experiment of both datasets. This is due to the efficiency of prediction system that accurately forecasts the future resource requirement. The overload prediction accuracy is approximately 99.94\% and 99.91 \% for GCD and PL, respectively. 

\subsection{Resource Scaling}
\subsubsection{Number of Active Servers}

\begin{figure}[!htbp]
	
	\centering
	\subfigure[Google Cluster ]{\includegraphics[width=.235\textwidth]{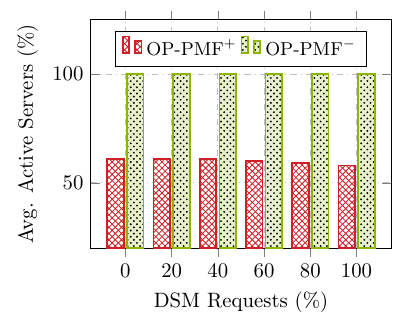}}
	\subfigure[Planet Lab ]{\includegraphics[width=.235\textwidth]{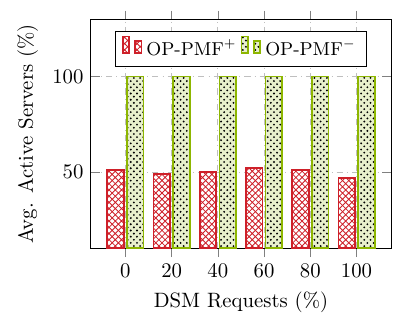}}\hfill
	
	\caption{Active Server Machines}
	\label{APM}
	
\end{figure}

\begin{figure}[!htbp]
	
	\centering
	\subfigure[Google Cluster ]{\includegraphics[width=.23\textwidth]{ 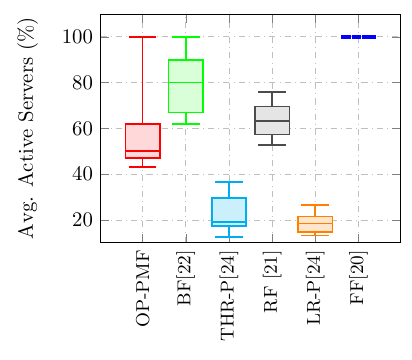}}
	\subfigure[Planet Lab ]{\includegraphics[width=.23\textwidth]{ 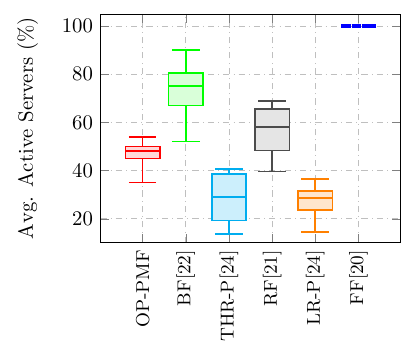}}\hfill
	
	\caption{Active Servers: OP-PMF  vs existing approaches}
	\label{ComparisonAPM}
	
\end{figure}

Fig. \ref{APM} compares the average percentage of total number of active servers versus different percentages of DSM requests $r^{DSM}(\%)$ by utilizing proposed framework (OP-PMF $^+$) and without proposed framework (OP-PMF$^-$). Various percentages of DSM requests $r^{DSM}(\%)$ represents different ratios of $r^{DSM}:r^{BEM}$ such as $r^{DSM} (\%)=20\%$ depicts $r^{DSM}:r^{BEM}= 1:4$. It is to be noted that OP-PMF$^+$ has reduced average percentage of total number of active servers up to $49\%$ and $51\%$ in case of GCD and PL, respectively because it estimates future resource usage of VMs and determines number and size of VMs to be requested in near future by applying proposed real-time resource prediction and VM scaling, respectively. Fig. \ref{ComparisonAPM} compares average percentage of total number of active servers of OP-PMF  with existing approaches: BF \cite{shirvastava2017best}, FF \cite{jangiti2019aggregated}, RF \cite{jung2010mistral}, THR-P \cite{hieu2017virtual}, and LR-P \cite{hieu2017virtual}, where OP-PMF shows stable performance with mean value of 50\% for VM traces of both workloads over existing approaches.  

\subsubsection{Resource Utilization} 
\begin{figure*}[!htbp]
	\centering	
	\subfigure[Google Cluster]{\includegraphics[width=0.3\linewidth, scale=2]{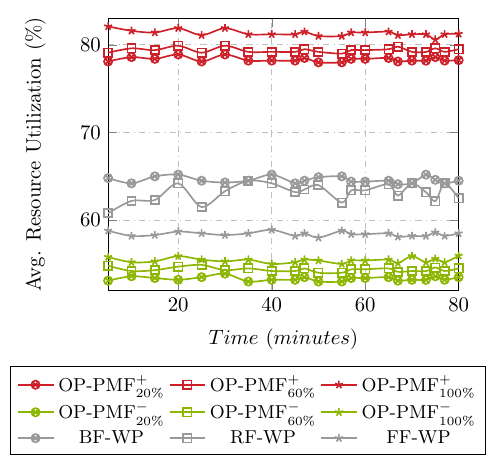}} \hfill
	\subfigure[Planet Lab]{\includegraphics[width=0.3\linewidth, scale=2]{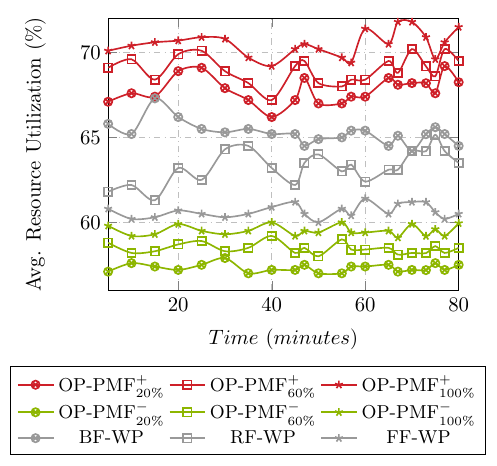}} \hfill
	\subfigure[OP-PMF vs existing approaches]{\includegraphics[width=0.28\linewidth, ]{ 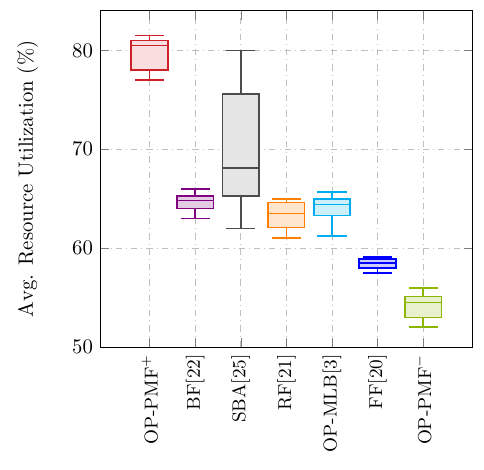}} 	
	\caption{Resource Utilization}
	\label{ResourceUtilization}
	
\end{figure*}

The resource utilization ($RU$) is evaluated by applying Eqs. (\ref{RU1}) and (\ref{RU2}), where $\mathds{N}$ is the number of resources, $RU^C$ and $RU^M$ are CPU and memory utilization of server. If $i^{th}$ physical machine ($S_i$) hosts VMs, it is active (i.e., $\alpha_i$ = 1), otherwise, inactive.  
\begin{equation}
RU^{\mathds{CDC}}= \frac{\sum_{i=1}^{P}{RU_i^{C}} + \sum_{i=1}^{P}{RU_i^{M}}}{|\mathds{N}| \times \sum_{i=1}^{P}{\alpha_i} } \label{RU1}
\end{equation}
\begin{equation}
RU_i^{{R}} = \frac{\sum_{j=1}^{Q}{\omega_{ji}} \times V_j^{{R}}}{S_i^{{R}}} \quad \forall_i \in \{1, P\}, {R} \in \{C, M\} \label{RU2}	
\end{equation} 
Figs. \ref{ResourceUtilization}(a) and \ref{ResourceUtilization}(b) compare average resource utilization percentage achieved by OP-PMF  with respect to various ratios of $r^{DSM}:r^{BEM}$ including 1:4 (i.e., OP-PMF$^+_{20\%}$), 3:2 (OP-PMF$^+_{60\%}$), and 1:0 (OP-PMF$^+_{100\%}$) against without OP-PMF framework (viz., OP-PMF$^-_{20\%}$, OP-PMF$^-_{60\%}$, and OP-PMF$^-_{100\%}$), Best Fit with VM prediction (BF-WP), Random Fit with VM prediction (RF-WP), and First Fit with VM prediction (FF-WP) for GCD and PL workloads, respectively. OP-PMF shows superior $RU$ by 60\%, 37.9\%, 21.25\%, and 18.75\% over OP-PMF$^-$, FF-WP, RF-WP, and BF-WP, respectively for GCD VM traces.
In case of PL VM traces, average $RU(\%)$ is improved by 60\%, 37.9\%, 21.25\%, and 18.75\% over OP-PMF$^-$, FF-WP, RF-WP, and BF-WP, respectively. Furthermore, Fig. \ref{ResourceUtilization}(c) compares box-plots for average $RU (\%)$ of OP-PMF against existing state-of-the-art approaches: SBA \cite{alanazi2017reducing}, OP-MLB \cite{saxena2021op}, BF \cite{shirvastava2017best}, and FF \cite{jangiti2019aggregated}, where it shows significant improvement in terms of median quartile up to 81\% for GCD VM traces. A boxplot distributes $RU (\%)$ statistically through quartiles where bottom, middle and top of the box are the first, second, and third quartiles, while the ends of the whiskers are the least and highest of all values, respectively.

\subsubsection{VM Scaling}
\begin{figure}[!htbp]
	\centering
	\includegraphics[width=0.6\linewidth]{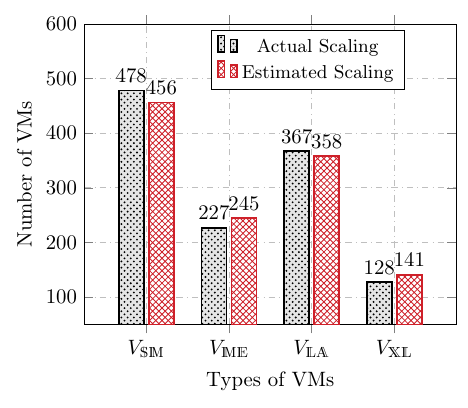}
	\caption{VM Scaling}
	\label{fig:vmscaling}
\end{figure}

The effect of self-adaptive scaling of VMs is analyzed in Fig. \ref{fig:vmscaling}, which reports the comparison of estimated and actual number of different VMs types scaled during experimental execution of different requests on VMs with resource utilization from GCD. It is to be noticed that proposed resource prediction and 
VM autoscaling approach has precisely determined the required number of different types of VMs (viz. $V_{\mathds{SM}}$, $V_{\mathds{ME}}$, $V_{\mathds{LA}}$,  $V_{\mathds{XL}}$) for execution of future requests beforehand. Therefore, OP-PMF  provides near-optimal VM autoscaling for $\mathds{CDC}$. Otherwise, SLA violations, performance degradation, and resource wastage may occur due to an inappropriate scaling of VMs.

\subsubsection{Power Consumption} 
\begin{figure*}[!htbp]
	\centering	
	\subfigure[Google Cluster]{\resizebox{0.3\textwidth}{!}{\begin{tikzpicture}
			\begin{axis}[
			width=.55\textwidth,
			height=0.45\textwidth,
			ybar,
			ymin=0,
			ymax=28,
			ybar=1.5pt,
			ymajorgrids=true,
			xmajorgrids=true,
			grid style=dashdotted,
			bar width=15pt,
			enlarge x limits=0.15,
			legend style={at={(0.5,0.95)},
				anchor=north,legend columns=-1},
			xlabel={DSM Requests (\%)},
			ylabel={Avg. PW Consumption ($KW$)}]
			\addplot[fill=white!20!,thick, pattern color=r4, draw=r4, postaction={
				pattern=crosshatch, thin}] coordinates{(0, 10.32) (20,9.929) (40,9.917) (60, 9.568) (80, 9.372) (100,9.214 )}; \addlegendentry{\small{OP-PMF$^+$}}
			\addplot[ fill=s3!20!,draw=s3, thick,pattern color=black, thick,  postaction={
				pattern=crosshatch dots
			}] coordinates{(0,19.56 ) (20, 19.07)(40,18.93) (60,18.9)  (80,18.58) (100,18.2 )}; \addlegendentry{\small{OP-PMF$^-$}}
			\end{axis}
			\end{tikzpicture}			
	}}		
	\subfigure[Planet Lab]{\resizebox{0.3\textwidth}{!}{
			\begin{tikzpicture}
			\begin{axis}[
			width=.55\textwidth,
			height=0.45\textwidth,
			ybar,
			ymin=3,
			ymax=23,
			ybar=1.5pt,
			ymajorgrids=true,
			xmajorgrids=true,
			grid style=dashdotted,
			bar width=15pt,
			enlarge x limits=0.15,
			legend style={at={(0.5,0.95)},
				anchor=north,legend columns=-1},
			xlabel={DSM Requests (\%)},
			ylabel={Avg. PW Consumption ($KW$)}]
			\addplot[fill=white!20!,thick, pattern color=r4, draw=r4, postaction={
				pattern=crosshatch, thin}] coordinates{(0, 7.708) (20,7.738) (40,8.853) (60, 7.883) (80, 7.764) (100,7.559 )}; \addlegendentry{\small{OP-PMF$^+$}}
			\addplot[ fill=s3!20!,draw=s3, thick,pattern color=black, thick,  postaction={
				pattern=crosshatch dots
			}] coordinates{(0,17.56 ) (20, 17.07)(40,17.93) (60,17.9)  (80,17.58) (100,17.2 )}; \addlegendentry{\small{OP-PMF$^-$}}
			
			\end{axis}
			\end{tikzpicture}
	}}		
	\subfigure[Comparative Analysis]{\includegraphics[width=0.26\linewidth]{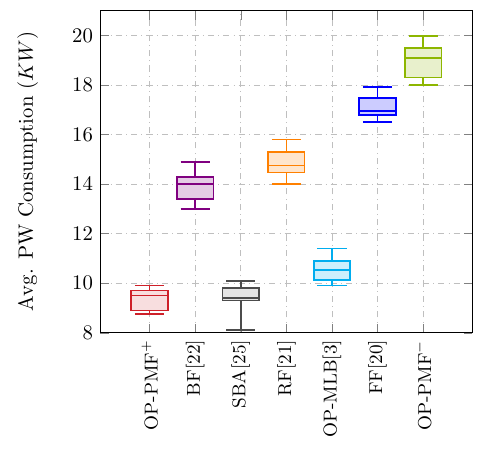}} 	
	\caption{Power Consumption}
	\label{power}
	
\end{figure*}

The power consumption of $\mathds{CDC}$ i.e., $PW^{\mathds{CDC}}$ is assessed by utilizing Eq. (\ref{power2}), where ${PW_i}^{Max}$, ${PW_i}^{Min}$, and ${PW_i}^{Idle}$ are maximum, minimum, and idle state power consumption, respectively of server ($S_i$). 
\begin{equation}
PW^{\mathds{CDC}} = 
\sum_{i=1}^{P} {({PW_i}^{Max} - {PW_i}^{Min})\times{RU} + {PW_i}^{Idle}}
\label{power2}
\end{equation}
The average $PW$ using OP-PMF$^+$ and OP-PMF$^-$ is compared for different ratios of  $r^{DSM}$ and $r^{BEM}$ reported in Fig. \ref{power}(a), where $PW$ is almost 50\% lesser for OP-PMF$^+$ as compared to OP-PMF$^-$ for GCD VM traces. Moreover,  Fig. \ref{power}(b) depicts that average $PW$ by OP-PMF$^+$ has reduced up to 60.7\% over OP-PMF$^-$ in case of PL VM traces. The comparative analysis of proposed framework and existing approaches: SBA \cite{alanazi2017reducing}, OP-MLB \cite{saxena2021op}, BF \cite{shirvastava2017best}, and FF \cite{jangiti2019aggregated} is shown in Fig. \ref{power}(c) using box-plots, where all the quartiles of OP-PMF$^+$ box-plot shows lesser values from the respective quartiles of comparative approaches except lower quartile of SBA box-plot. Therefore, OP-PMF substantially reduces $PW$ over existing approaches.
\subsubsection{ Electricity Cost} 
\begin{figure*}[!htbp]
	
	\centering
	\subfigure[Google Cluster ]{\resizebox{0.3\textwidth}{!}{\begin{tikzpicture}
			\begin{axis}[
			width=.55\textwidth,
			height=0.45\textwidth,
			ybar,
			ymin=0.2,
			ymax=2,
			ybar=1.5pt,
			ymajorgrids=true,
			xmajorgrids=true,
			grid style=dashdotted,
			bar width=15pt,
			enlarge x limits=0.15,
			legend style={at={(0.5,0.95)},
				anchor=north,legend columns=-1},
			xlabel={\small{DSM Requests (\%)}},
			ylabel={\small{Avg. Electricity Cost/Hour (\$)}}]
			\addplot[fill=white!20!,thick, pattern color=r4, draw=r4, postaction={
				pattern=crosshatch, thin}] coordinates{(0, 7.708*0.0875) (20,7.738*0.0875-0.0208*0.2) (40,8.853*0.0875-0.0208*0.4) (60, 7.883*0.0875-0.0208*0.6) (80, 7.764*0.0875-0.0208*0.8) (100,7.559*0.0875 -0.0208)}; \addlegendentry{\small{OP-PMF$^+$}}
			\addplot[ fill=s3!20!,draw=s3, thick,pattern color=black, thick,  postaction={
				pattern=crosshatch dots
			}] coordinates{(0,17.56*0.0875 ) (20, 17.07*0.0875-0.0208*0.2)(40,17.93*0.0875-0.0208*0.4) (60,17.9*0.0875-0.0208*0.6)  (80,17.58*0.0875-0.0208*0.8) (100,17.2*0.0875-0.0208 )}; \addlegendentry{\small{OP-PMF$^-$}}
			
			\end{axis}
			\end{tikzpicture}}}
	\subfigure[Planet Lab ]{\resizebox{0.3\textwidth}{!}{\begin{tikzpicture}
			\begin{axis}[
			width=.55\textwidth,
			height=0.45\textwidth,
			ybar,
			ymin=0.5,
			ymax=2.2,
			ybar=1.5pt,
			ymajorgrids=true,
			xmajorgrids=true,
			grid style=dashdotted,
			bar width=15pt,
			enlarge x limits=0.15,
			legend style={at={(0.5,0.95)},
				anchor=north,legend columns=-1},
			xlabel={\small{DSM Requests (\%)}},
			ylabel={\small{Avg. Electricity Cost/Hour (\$)}}]
			\addplot[fill=white!20!,thick, pattern color=r4, draw=r4, postaction={
				pattern=crosshatch, thin}] coordinates{(0, 10.32*0.0875) (20,9.929*0.0875-0.0208*0.2) (40,9.917*0.08750-.0208*0.4) (60, 9.568*0.08750-.0208*0.6) (80, 9.372*0.0875-.0208*0.8) (100,9.214*0.0875 -.0208*1)}; \addlegendentry{\small{OP-PMF$^+$}}
			\addplot[ fill=s3!20!,draw=s3, thick,pattern color=black, thick,  postaction={
				pattern=crosshatch dots
			}] coordinates{(0,19.56 *0.0875) (20, 19.07*0.0875-.0208*0.2)(40,18.93*0.0875-.0208*0.4) (60,18.9*0.0875-.0208*0.6)  (80,18.58*0.0875-.0208*0.8) (100,18.2*0.0875-.0208*1 )}; \addlegendentry{\small{OP-PMF$^-$}}				
			\end{axis}
			\end{tikzpicture}		
	}}
	\subfigure[Cost Comparison]{\includegraphics[width=0.26\linewidth ]{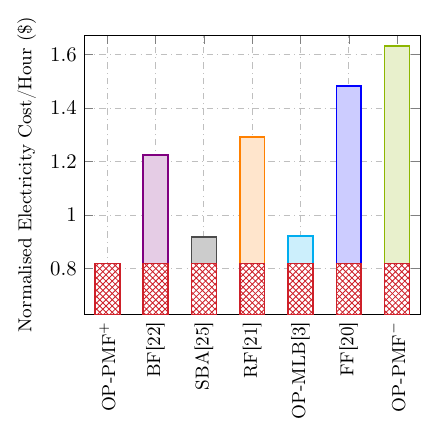}} 
	\caption{Electricity Bill Computation}
	\label{Bill}
	
\end{figure*}

The electricity cost expenditure ($\mathds{EC}$) is estimated using $\mathds{CDC}$ electricity price values equals to $0.07 \$/KWH$, which is obtained on the basis of average electricity price in U.S. \cite{usEnergy2014}.  The extra charge for earlier execution of $r^{DSM}$ is $0.25 \$/H$. Figs. \ref{Bill}(a) and \ref{Bill}(b) show comparison of average electricity cost consumed per hour with OP-PMF$^+$ and OP-PMF$^-$ for GCD and PL VM traces, respectively. It is observed that average $\mathds{EC}$ with OP-PMF$^+$ is reduced up to $53.3\%$ and $55.56\%$ over OP-PMF$^-$ in case of $100\%$ $r^{DSM}$ for GCD and PL VM traces, respectively. Furthermore, total electricity expenditure for execution of GCD VM traces using OP-PMF and other state-of-the-art approaches normalised with respect to OP-PMF electricity cost expenditure is shown in Fig. \ref{Bill}(c). The comparison of $\mathds{EC}$ follows the order: OP-PMF$^+$ $<$ SBA \cite{alanazi2017reducing} $\simeq$ OP-MLB \cite{saxena2021op} $<$ BF-WP \cite{shirvastava2017best} $<$ RF-WP \cite{jung2010mistral} $<$FF-WP \cite{jangiti2019aggregated}$<$ OP-PMF$^-$.           

\subsubsection{Earned Profits} 
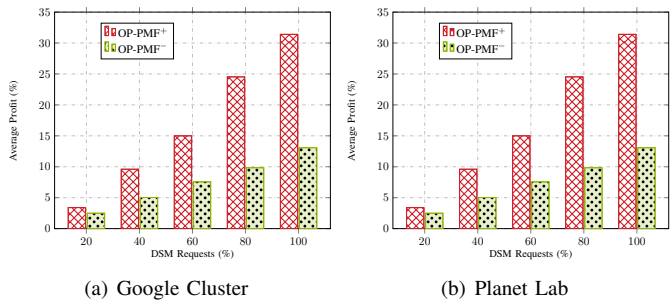
\begin{figure}[!htbp]
	
	\centering
	\subfigure[Google Cluster ]{\resizebox{0.24\textwidth}{!}{\begin{tikzpicture}
			\begin{axis}[
			width=.55\textwidth,
			height=0.45\textwidth,
			ybar,
			ymin=0,
			ymax=35,
			ybar=1.5pt,
			ymajorgrids=true,
			xmajorgrids=true,
			grid style=dashdotted,
			bar width=15pt,
			enlarge x limits=0.15,
			legend style={at={(0.3,0.95)},
				anchor=north,legend columns=1},
			xlabel={\small{DSM Requests (\%)}},
			ylabel={\small{Average Profit (\%)}}]
			\addplot[fill=white!20!,thick, pattern color=r4, draw=r4, postaction={
				pattern=crosshatch, thin}] coordinates{(20,3.4) (40,0.96*10) (60, 1.5*10) (80, 24.5293) (100,31.4)}; \addlegendentry{OP-PMF$^+$}
			\addplot[ fill=s3!20!,draw=s3, thick,pattern color=black, thick,  postaction={
				pattern=crosshatch dots
			}] coordinates{(20, 2.49)(40,5.03048) (60,7.557)  (80,9.841) (100,13.08)}; \addlegendentry{OP-PMF$^-$}
			
			\end{axis}
			\end{tikzpicture}
			
	}}
	\subfigure[Planet Lab ]{\resizebox{0.24\textwidth}{!}{\begin{tikzpicture}
			\begin{axis}[
			width=.55\textwidth,
			height=0.45\textwidth,
			ybar,
			ymin=0,
			ymax=35,
			ybar=1.5pt,
			ymajorgrids=true,
			xmajorgrids=true,
			grid style=dashdotted,
			bar width=15pt,
			enlarge x limits=0.15,
			legend style={at={(0.3,0.95)},
				anchor=north,legend columns=1},
			xlabel={\small{DSM Requests (\%)}},
			ylabel={\small{Average Profit (\%)}}]
			\addplot[fill=white!20!,thick, pattern color=r4, draw=r4, postaction={
				pattern=crosshatch, thin}] coordinates{(20,3.4) (40,0.96*10) (60, 1.5*10) (80, 24.5293) (100,31.4)}; \addlegendentry{OP-PMF$^+$}
			\addplot[ fill=s3!20!,draw=s3, thick,pattern color=black, thick,  postaction={
				pattern=crosshatch dots
			}] coordinates{(20, 2.49)(40,5.03048) (60,7.557)  (80,9.841) (100,13.08)}; \addlegendentry{OP-PMF$^-$}
			
			\end{axis}
			\end{tikzpicture}
			
	}}	
	\caption{Earned Profit of Cloud Data Centre}
	\label{Profit}
	
\end{figure}

The total profit is estimated in terms of reduction in electricity cost per hour with respect to varying percentage of $r^{DSM}$ from $20\%$ to $100\%$ as follows: $r^{DSM} (\%) \times 0.25 \times te$, where $te$ is time of earlier execution which is considered equals to 5 minutes (as per the time-interval of collection of VM traces in both datasets) for $r^{DSM}$. The same procedure is followed to compute profit earned with OP-PMF$^+$ and OP-PMF$^-$ approaches. Figs. \ref{Profit}(a) and \ref{Profit}(b) report that OP-PMF$^+$ always earns more profit as compared to OP-PMF$^-$ which increases with percentage of $r^{DSM}$ achieving maximum average profit up to $49.72\%$ and $51.18\%$ for GCD and PL VM traces, respectively. These results are attributed to the fact that OP-PMF$^+$ has much lesser $PW$ and $\mathds{EC}$ than OP-PMF$^-$, while the gained reduction in electricity bill is equal for both approaches. Though, for the experiments, deadline of request completion and time of earlier execution are fixed (as these values are not available in original datasets), OP-PMF framework promises high potential to enhance profits and reduce electricity bills for $\mathds{CDC}$ when utilized with actual values of request deadline and earlier execution time.

\subsection{Trade-off between Earned Profits and Power Consumption}
\begin{figure} [!htbp]
    \centering
    \includegraphics[width=0.65\linewidth, scale=2]{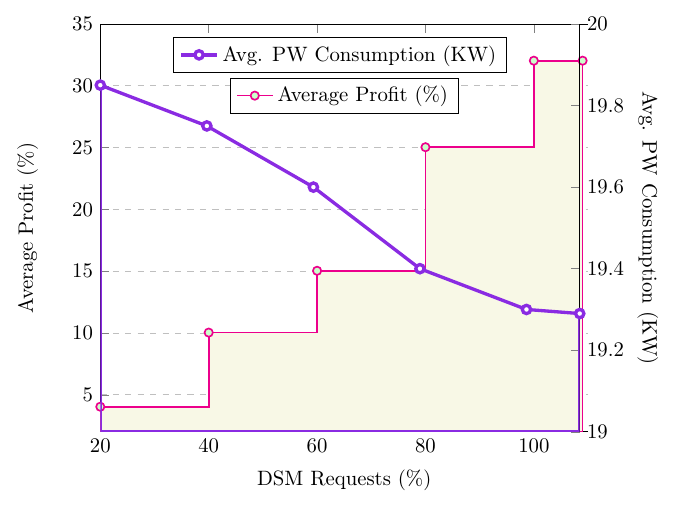}
    \caption{Google Cluster: Profit v/s power consumption}
    \label{fig:GCD_comparison.pdf}
\end{figure}

\begin{figure} [!htbp]
    \centering
    \includegraphics[width=0.65\linewidth, scale=2]{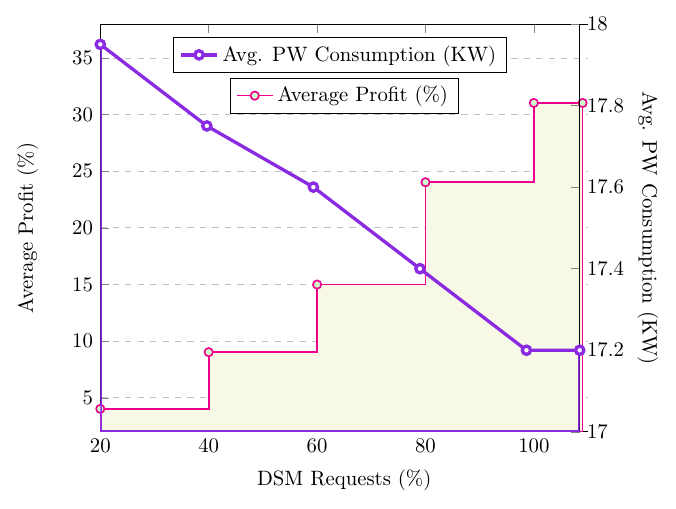}
    \caption{Planet Lab: Profit v/s power consumption}
    \label{fig:PL_comparison.pdf}
\end{figure}

The trade-off and the inherent  balance between maximizing profit and minimizing power consumption for GCD VMs traces and PL VMs traces is reported in Fig. \ref{fig:GCD_comparison.pdf} and Fig. \ref{fig:PL_comparison.pdf}, respectively.  This observation demonstrates that optimizing resource usage and power consumption, alongside increased execution of requests with DSM service pricing, significantly boosts profit for both data traces. Consequently, the proposed OP-PMF framework offers pragmatic industry cloud resource management by effectively leveraging service pricing for profit maximization.

\section{Conclusions and Future work} \label{sec-conclusion}

This paper proposed OP-PMF, a novel framework for industry cloud resource management that harnesses resource oversubscription and heterogeneous service pricing models. Its objective is to maximize profits while minimizing operational costs, including electricity expenditure, active servers, VM migrations, resource wastage, and power consumption. OP-PMF integrates an adaptive ensemble machine learning-driven prediction model to proactively estimate VM resource utilization and optimize VM deployment before executing user requests. Moreover, it introduces two cloud service pricing models, the Delay Sensitive Model and the Best-Effort Model, to classify and execute user requests effectively. OP-PMF is evaluated through experimental simulations and comparison with state-of-the-art approaches using benchmark VM traces. The results demonstrate significant improvements in electricity bill reduction, power consumption, number of active servers, resource utilization, and earned profits. The future work aims to extend OP-PMF to handle VM failovers effectively, ensuring seamless operation and continuity in the event of VM failures. A robust failover mechanisms can be incorporated to further enhance the reliability and resilience of cloud services, contributing to improved user experience and performance. 





\bibliographystyle{IEEEtran}
\bibliography{bibfile}

\begin{thebibliography}{10}
\providecommand{\url}[1]{#1}
\csname url@samestyle\endcsname
\providecommand{\newblock}{\relax}
\providecommand{\bibinfo}[2]{#2}
\providecommand{\BIBentrySTDinterwordspacing}{\spaceskip=0pt\relax}
\providecommand{\BIBentryALTinterwordstretchfactor}{4}
\providecommand{\BIBentryALTinterwordspacing}{\spaceskip=\fontdimen2\font plus
\BIBentryALTinterwordstretchfactor\fontdimen3\font minus
  \fontdimen4\font\relax}
\providecommand{\BIBforeignlanguage}[2]{{%
\expandafter\ifx\csname l@#1\endcsname\relax
\typeout{** WARNING: IEEEtran.bst: No hyphenation pattern has been}%
\typeout{** loaded for the language `#1'. Using the pattern for}%
\typeout{** the default language instead.}%
\else
\language=\csname l@#1\endcsname
\fi
#2}}
\providecommand{\BIBdecl}{\relax}
\BIBdecl

\bibitem{datacenterdynamics2021}
D.~Dynamics, ``Data centers environment 2021 report state green data center,''
  \emph{Available:
  https://www.datacenterdynamics.com/en/whitepapers/data-centers-environment-2021},
  2021.

\bibitem{kaur2019big}
K.~Kaur, S.~Garg, G.~Kaddoum, E.~Bou-Harb, and K.-K.~R. Choo, ``A big
  data-enabled consolidated framework for energy efficient software defined
  data centers in iot setups,'' \emph{IEEE Transactions on Industrial
  Informatics}, vol.~16, no.~4, pp. 2687--2697, 2019.

\bibitem{saxena2021op}
D.~Saxena, A.~K. Singh, and R.~Buyya, ``\uppercase{OP-MLB}: An online vm
  prediction based multi-objective load balancing framework for resource
  management at cloud datacenter,'' \emph{IEEE Transactions on Cloud
  Computing}, vol. doi: 10.1109/TCC.2021.3059096, 2021.

\bibitem{saxena2022high}
D.~Saxena and A.~K. Singh, ``A high availability management model based on vm
  significance ranking and resource estimation for cloud applications,''
  \emph{IEEE Transactions on Services Computing}, 2022.

\bibitem{li2019transforming}
Y.~Li, Y.~Wen, D.~Tao, and K.~Guan, ``Transforming cooling optimization for
  green data center via deep reinforcement learning,'' \emph{IEEE transactions
  on cybernetics}, vol.~50, no.~5, pp. 2002--2013, 2019.

\bibitem{yuan2016ttsa}
H.~Yuan, J.~Bi, W.~Tan, M.~Zhou, B.~H. Li, and J.~Li, ``Ttsa: An effective
  scheduling approach for delay bounded tasks in hybrid clouds,'' \emph{IEEE
  transactions on cybernetics}, vol.~47, no.~11, pp. 3658--3668, 2016.

\bibitem{singh2015qos}
S.~Singh and I.~Chana, ``Qos-aware autonomic resource management in cloud
  computing: a systematic review,'' \emph{ACM Computing Surveys (CSUR)},
  vol.~48, no.~3, pp. 1--46, 2015.

\bibitem{saxena2023ai}
D.~Saxena, I.~Gupta, R.~Gupta, A.~K. Singh, and X.~Wen, ``An ai-driven vm
  threat prediction model for multi-risks analysis-based cloud cybersecurity,''
  \emph{IEEE Transactions on Systems, Man, and Cybernetics: Systems}, 2023.

\bibitem{singh2021quantum}
A.~K. Singh, D.~Saxena, J.~Kumar, and V.~Gupta, ``A quantum approach towards
  the adaptive prediction of cloud workloads,'' \emph{IEEE Transactions on
  Parallel and Distributed Systems}, 2021.

\bibitem{saxena2024high}
D.~Saxena and A.~K. Singh, ``A high up-time and security centered resource
  provisioning model towards sustainable cloud service management,'' \emph{IEEE
  Transactions on Green Communications and Networking}, 2024.

\bibitem{singh2023bio}
A.~K. Singh, S.~R. Swain, D.~Saxena, and C.-N. Lee, ``A bio-inspired virtual
  machine placement toward sustainable cloud resource management,'' \emph{IEEE
  Systems Journal}, 2023.

\bibitem{son2017sla}
J.~Son, A.~V. Dastjerdi, R.~N. Calheiros, and R.~Buyya, ``Sla-aware and
  energy-efficient dynamic overbooking in sdn-based cloud data centers,''
  \emph{IEEE Transactions on Sustainable Computing}, vol.~2, no.~2, pp. 76--89,
  2017.

\bibitem{molto2016automatic}
G.~Molt{\'o}, M.~Caballer, and C.~De~Alfonso, ``Automatic memory-based vertical
  elasticity and oversubscription on cloud platforms,'' \emph{Future Generation
  Computer Systems}, vol.~56, pp. 1--10, 2016.

\bibitem{reiss2011google}
C.~Reiss, J.~Wilkes, and J.~L. Hellerstein, ``Google cluster-usage traces:
  format+ schema,'' \emph{Google Inc., White Paper}, pp. 1--14, 2011.

\bibitem{sharma2016multi}
N.~K. Sharma and G.~R.~M. Reddy, ``Multi-objective energy efficient virtual
  machines allocation at the cloud data center,'' \emph{IEEE Trans. on Serv.
  Comput.}, vol.~12, no.~1, pp. 158--171, 2016.

\bibitem{minas2009energy}
L.~Minas and B.~Ellison, \emph{Energy efficiency for information technology:
  How to reduce power consumption in servers and data centers}.\hskip 1em plus
  0.5em minus 0.4em\relax Intel Press, 2009.

\bibitem{Dell1999}
Dell, ``Power model. [online].'' \emph{https://
  www.dell.com/systems/power/hardware/}, 1999.

\bibitem{usEnergy2014}
U.~E.~I. Administration, ``U.s. electricity prices.''
  \emph{[Online].Available:http://www.eia.gov}, 2014.

\bibitem{pinciroli2020cedule+}
R.~Pinciroli, A.~Ali, F.~Yan, and E.~Smirni, ``Cedule+: Resource management for
  burstable cloud instances using predictive analytics,'' \emph{IEEE
  Transactions on Network and Service Management}, 2020.

\bibitem{jangiti2019aggregated}
S.~Jangiti, E.~S. Ram, and V.~S. Sriram, ``Aggregated rank in
  first-fit-decreasing for green cloud computing,'' in \emph{Cognitive
  Informatics and Soft Computing}.\hskip 1em plus 0.5em minus 0.4em\relax
  Springer, 2019, pp. 545--555.

\bibitem{jung2010mistral}
G.~Jung, M.~A. Hiltunen, K.~R. Joshi, R.~D. Schlichting, and C.~Pu, ``Mistral:
  Dynamically managing power, performance, and adaptation cost in cloud
  infrastructures,'' in \emph{2010 IEEE 30th Int'l Conf. on Distributed
  Computing Systems}.\hskip 1em plus 0.5em minus 0.4em\relax IEEE, 2010, pp.
  62--73.

\bibitem{shirvastava2017best}
S.~Shirvastava, R.~Dubey, and M.~Shrivastava, ``Best fit based vm allocation
  for cloud resource allocation,'' \emph{International Journal of Computer
  Applications}, vol. 158, no.~9, 2017.

\bibitem{dabbagh2015exploiting}
M.~Dabbagh, B.~Hamdaoui, M.~Guizani, and A.~Rayes, ``Exploiting task elasticity
  and price heterogeneity for maximizing cloud computing profits,'' \emph{IEEE
  Transactions on Emerging Topics in Computing}, vol.~6, no.~1, pp. 85--96,
  2015.

\bibitem{hieu2017virtual}
N.~T. Hieu, M.~Di~Francesco, and A.~Yl{\"a}-J{\"a}{\"a}ski, ``Virtual machine
  consolidation with multiple usage prediction for energy-efficient cloud data
  centers,'' \emph{IEEE Transactions on Services Computing}, vol.~13, no.~1,
  pp. 186--199, 2017.

\bibitem{alanazi2017reducing}
S.~Alanazi, M.~Dabbagh, B.~Hamdaoui, M.~Guizani, and N.~Zorba, ``Reducing data
  center energy consumption through peak shaving and locked-in energy
  avoidance,'' \emph{IEEE Transactions on Green Communications and Networking},
  vol.~1, no.~4, pp. 551--562, 2017.

\bibitem{kumar2018renewable}
N.~Kumar, G.~S. Aujla, S.~Garg, K.~Kaur, R.~Ranjan, and S.~K. Garg, ``Renewable
  energy-based multi-indexed job classification and container management scheme
  for sustainability of cloud data centers,'' \emph{IEEE Transactions on
  Industrial Informatics}, vol.~15, no.~5, pp. 2947--2957, 2018.

\bibitem{sahoo2017lvrm}
P.~K. Sahoo, C.~K. Dehury, and B.~Veeravalli, ``Lvrm: On the design of
  efficient link based virtual resource management algorithm for cloud
  platforms,'' \emph{IEEE Transactions on Parallel and Distributed Systems},
  vol.~29, no.~4, pp. 887--900, 2017.

\bibitem{IBM1999}
IBM, ``Power model. [online].'' \emph{https:// www.ibm.com/}, 1999.

\bibitem{amazon1999EC2}
Amazon, ``Amazon ec2 instances. [online].'' \emph{https://
  aws.amazon.com/ec2/instance-types/}, 1999.

\bibitem{beloglazov2012optimal}
A.~Beloglazov and R.~Buyya, ``Optimal online deterministic algorithms and
  adaptive heuristics for energy and performance efficient dynamic
  consolidation of virtual machines in cloud data centers,'' \emph{Concurrency
  and Computation: Practice and Experience}, vol.~24, no.~13, pp. 1397--1420,
  2012.

\bibitem{dabbagh2016energy}
M.~Dabbagh, B.~Hamdaoui, M.~Guizani, and A.~Rayes, ``An energy-efficient vm
  prediction and migration framework for overcommitted clouds,'' \emph{IEEE
  Transactions on Cloud Computing}, vol.~6, no.~4, pp. 955--966, 2018.

\end{thebibliography}
\begin{IEEEbiography}[{\includegraphics[width=0.8\linewidth]{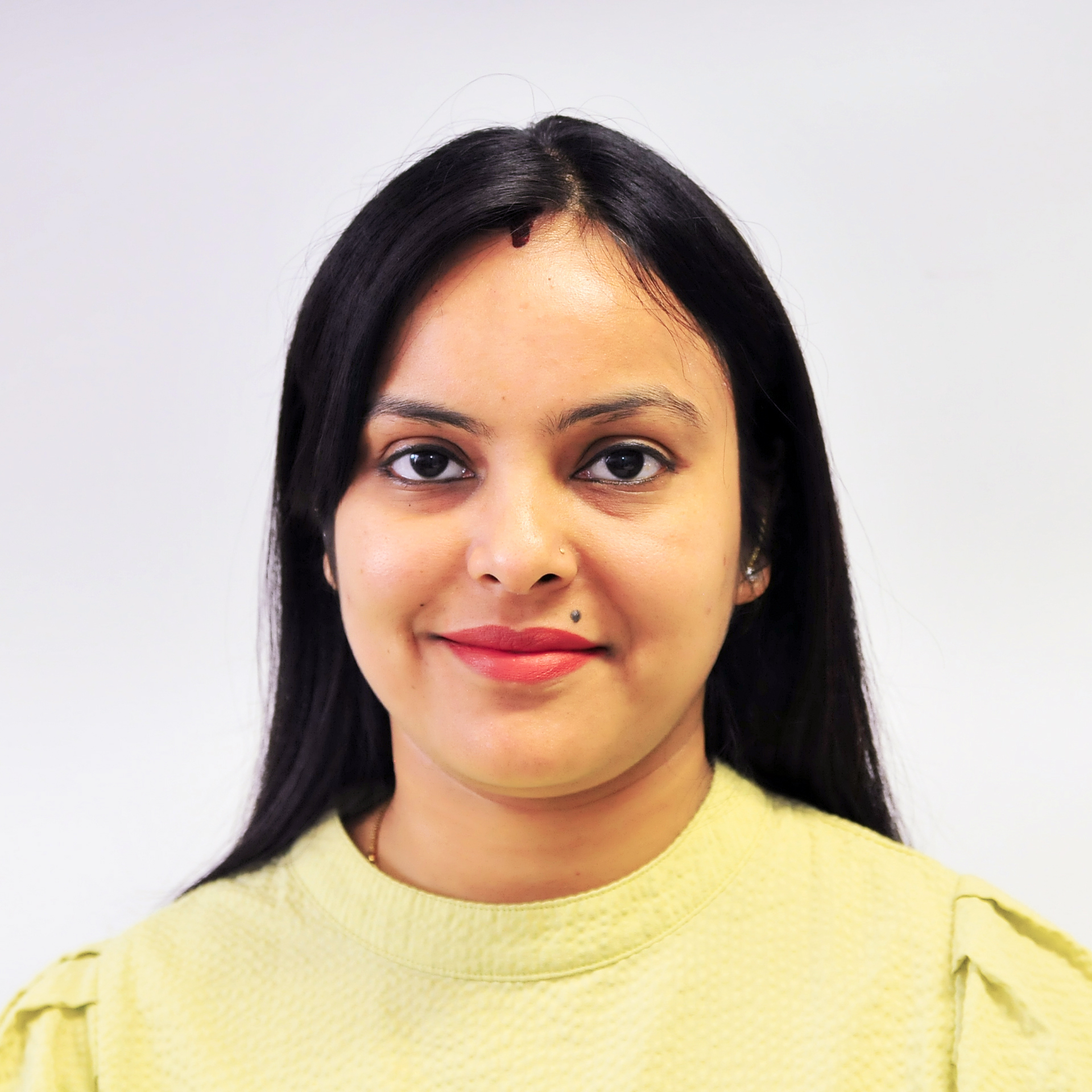}}]{Deepika Saxena} is working as an Associate Professor in the Division of Information Systems at the University of Aizu, Japan. She earned her Ph.D. degree in Computer Science from the National Institute of Technology, Kurukshetra, India, and completed her Post Doctorate from the Department of Computer Science at Goethe University, Frankfurt, Germany.  She is the recipient of the prestigious IEEE TCSC 2023 Outstanding Ph.D. Dissertation Award and EUROSIM 2023 Best Ph.D. Thesis Award. She is the recipient of  the prestigious Japan Society for the Promotion of Science (JSPS) KAKENHI Early Career Young Scientist Research Grant FY2024. Also, her research paper, published in the IEEE Transactions on Cloud Computing Journal, received the 2022 Best Paper Award from the IEEE Computer Society Publications Board. Her major research interests include Neural networks, Evolutionary algorithms, Resource management and Security in Cloud Computing, Internet traffic management, and Quantum machine learning, DataLakes, Dynamic Caching Management.
\end{IEEEbiography}

\begin{IEEEbiography}[{\includegraphics[width=0.8\linewidth]{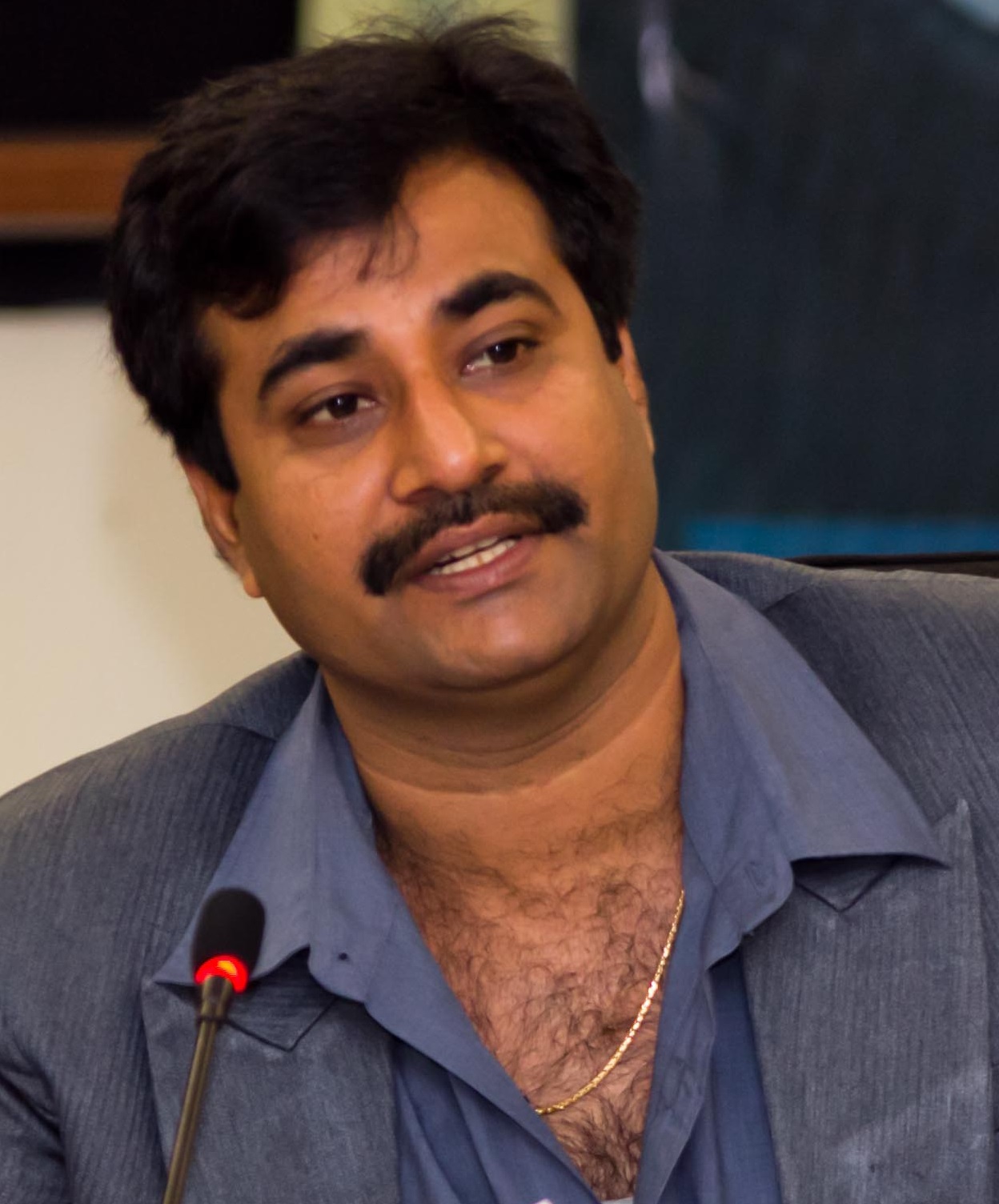}}]{Ashutosh Kumar Singh}
	 is currently serving as the Director and Professor at the Indian Institute of Information Technology Bhopal. Also, he is working as an Adjunct Professor in the University of Economics and Human Sciences, Warsaw, Poland. He received his Ph.D. in Electronics Engineering from Indian Institute of Technology, BHU, India and Post Doc from the Department of Computer Science, University of Bristol, UK. He has research and teaching experience in various Universities of India, the UK, and Malaysia. His research area includes the Design and Testing of Digital Circuits, Data Science, Cloud Computing, Machine Learning, and Security. He has published more than 400 research papers in different journals and conferences of high repute.  His research paper, published in the IEEE Transactions on Cloud Computing Journal, was honored with the 2022 Best Paper Award by the IEEE Computer Society Publications Board. 
	
\end{IEEEbiography}
\end{document}